\documentclass[journal=jpcc,manuscript=article]{achemso}
\usepackage{color}

\definecolor{red}{rgb}{0.8, 0.0, 0.0}
\definecolor{blue}{rgb}{0.06, 0.2, 0.65}
\definecolor{green}{rgb}{0.0, 0.6, 0.0}
\usepackage{multirow}
\newcommand{\onlinecite}[1]{\hspace{-1 ex} \nocite{#1}\citenum{#1}} 
\usepackage{hyperref}
\usepackage{fdsymbol}
\author{Melisa M. Gianetti}
\affiliation{Dipartimento di Fisica, Università degli Studi di Milano, Via Celoria 16, Milano, 20133 Italy}
\email{melisamariel@gmail.com}
\author{Roberto Guerra}
\affiliation{Center for Complexity and Biosystems, Department of Physics, University of Milan, via Celoria 16, Milano, 20133, Italy}
\author{Andrea Vanossi}
\affiliation{CNR-IOM, Consiglio Nazionale delle Ricerche - Istituto Officina dei Materiali, c/o SISSA, Via Bonomea 265, 34136 Trieste, Italy}%
\alsoaffiliation{International School for Advanced Studies (SISSA), Via Bonomea 265, 34136 Trieste, Italy\\}%
\author{Michael Urbakh}
 \affiliation{Department of Physical Chemistry, School of Chemistry, The Raymond and Beverly Sackler Faculty of Exact Sciences and The Sackler Center for Computational Molecular and Materials Science, Tel Aviv University, Tel Aviv 6997801, Israel}%
\author{Nicola Manini}
\affiliation{Dipartimento di Fisica, Università degli Studi di Milano, Via Celoria 16, Milano, 20133 Italy}%

\title[An \textsf{achemso} demo]
  {Thermal Friction Enhancement in Zwitterionic Monolayers}


\begin{document}

\begin{tocentry}

\includegraphics[width=0.6\textwidth]{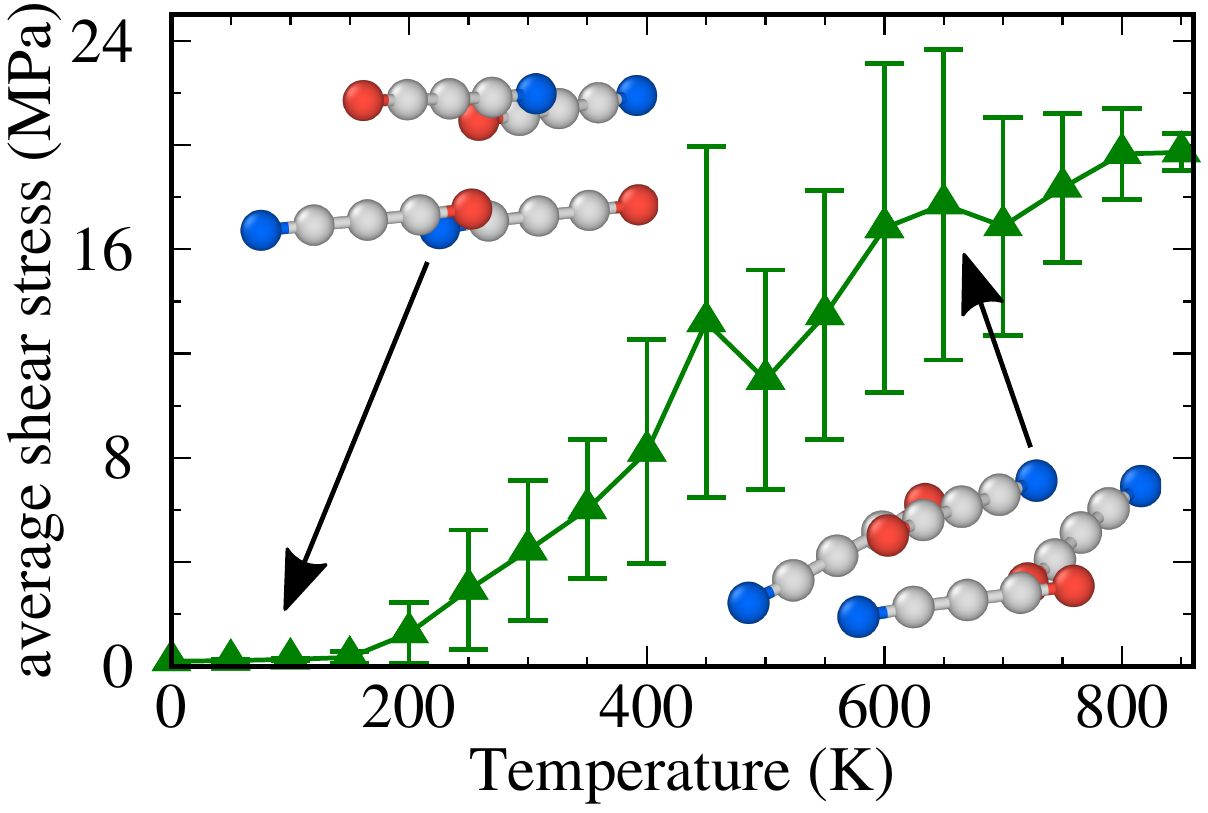}

%Some journals require a graphical entry for the Table of Contents.
%This should be laid out ``print ready'' so that the sizing of the

%Inside the \texttt{tocentry} environment, the font used is Helvetica
%8\,pt, as required by \emph{Journal of the American Chemical

%The surrounding frame is 9\,cm by 3.5\,cm, which is the maximum
%permitted for  \emph{Journal of the American Chemical Society}
%graphical table of content entries. The box will not resize if the
%content is too big: instead it will overflow the edge of the box.

%This box and the associated title will always be printed on a
%separate page at the end of the document.

\end{tocentry}

\begin{abstract}

We introduce a model for zwitterionic monolayers %charged polymer brushes 
and investigate its tribological response to changes in applied load, sliding velocity, and temperature by means of molecular-dynamics simulations. The proposed model exhibits different regimes of motion depending on temperature and sliding velocity. We find a remarkable increase of friction with temperature, which we attribute to the formation and rupture of transient bonds between individual %polymers
molecules 
of opposite sliding layers, triggered by the out-of-plane thermal fluctuations of the %polymers 
molecules' 
orientations. To highlight the effect of the molecular charges, we compare these results with analogous simulations for the %uncharged 
charge-free  
system. These findings are expected to be relevant to nanoscale rheology and tribology experiments of locally-charged lubricated systems such as, e.g., 
experiments performed on zwitterionic monolayers, phospholipid micelles, or
confined polymeric brushes in a surface force apparatus.

\end{abstract}

\section{\label{sec:intro}Introduction}

The possibility of controlling nano- and mesoscale friction and mechanical response in a variety of diverse physical systems has been investigated extensively in recent years~\cite{wu2015,Manini17,Vanossi18,Krim19}.
%
% examples of regular thermolubric behavior:
In particular, in confined geometries, friction is affected by temperature, usually exhibiting a regular ``thermolubric'' behavior, with friction decreasing 
as temperature increases at microscopic scales \cite{Sang01,Dudko02,Szlufarska08,Brukman08,Steiner09}.
The rationale for
this standard behavior
is random thermal fluctuations assisting the sliding interface in the negotiation of interlocking barriers, thus promoting advancement. 

The reverse, namely, friction increasing with temperature, is far less common, although it has been observed in specific
situations.~\cite{Schirmeisen06,Barel10a,Barel10b,sheng2012electrorheological,Vanossi18,zhu2020nano} 
Certainly, inverted thermolubricity in poor heat-transfer conditions may promote instabilities in the frictional dynamics:
the heat dissipated by friction itself can raise temperature, thus triggering a further increase in friction, eventually possibly leading to some kind of lockup, which cuts off this runaway condition.

% one more example from geophysics (add if relevant)
%For example, inverted thermolubricity has been explored through rate-and-state models in geophysical contexts~\cite{Singh2016}.

Inverted thermolubricity was considered primarily as an ingredient for phenomenological models \cite{Singh2016,Tian2018}, but it was also investigated in atomic-scale friction within the most basic and fundamental model, namely, the Prandtl-Tomlinson (PT) model \cite{Tshiprut09}, where it was shown 
that a peak in friction may
arise in a range of temperatures corresponding to a transition from a multiple-slip 
regime (low $T$) to a single-slip regime (high $T$).
However, that simple model fails to reproduce the observed features of the temperature and velocity %dependencies 
dependence of friction and of the corresponding force traces measured via atomic-force microscopy (AFM).

Modeling via molecular dynamics (MD) simulations, as a sort of controlled computational ``experiment", has been revealed to be extremely useful in investigating frictional processes of complex systems \cite{VanossiRMP13,Manini15,Manini16}, possibly avoiding interpretative pitfalls arising from indirect or ex-situ characterization of contact surfaces. 

In this work, we 
investigate the possibility of 
an inverted thermal dependence of friction % in  %``brushes'' of polymers formed by charge-carrying units confined between surfaces in relative motion.
%molecules containing 
between zwitterionic head groups 
%confined between surfaces
in relative sliding motion.
We 
investigate if, and how, the thermal disordering and rearrangement of such %brushes 
zwitterionic head groups
leads to an increase of friction, at least over suitable temperature ranges,  
especially those 
experimentally relevant.

%Polymer brushes are 
We simulate zwitterionic molecules,
flexible linear macromolecules that 
can be
tethered to a surface with the aim of modifying 
its 
distinctive properties. 
These %brushes 
molecules can have charge-free or %charged 
zwitterionic
terminations, depending on the specific surface features
that are addressed \cite{perkin2013}.
Applications involving %polymer brushes 
zwitterionic molecules 
include colloid stabilization, regulation in wetting and adhesion, 
and the formation 
of protective coatings, among many others \cite{ma2019brushing,myshkin2009adhesion,chen2009}.
Even though vast theoretical research on %polymer brush 
zwitterionic molecule 
lubrication has been carried out \cite{kreer2016polymer,debeer2013,deBeer2014}, there 
remain 
unanswered questions about the microscopic mechanisms of friction, 
especially under the influence of temperature, on surfaces decorated or covered with these complex molecules.

Here, we develop a coarse-grained model to study friction 
between two preassembled %polymer brushes 
zwitterionic monolayers 
\cite{raviv2003,yu2012langmuir,ron2014}.
Our simulations 
demonstrate
how the 
modification of the 
geometric rearrangement of locally %charged (zwitterionic) 
zwitterionic
molecular portions 
gives rise to different 
interlocking 
configurations
at the sliding interface, leading to distinct frictional regimes, as a function of temperature.

%While such experimental setup does measure the system rheological and dissipative response in terms of crucial, yet averaged, physical quantities (Carpick,Salmeron, Chem Rev 97, 1163 (1997))), the full exploitation of this electrotunable approach requires the necessity of casting light on the elemental mechanisms and molecular rearrangements occurring at the sheared interface.

\section{\label{sec:methods}Methods}

\begin{figure*}[tb!]
\includegraphics[width=\textwidth]{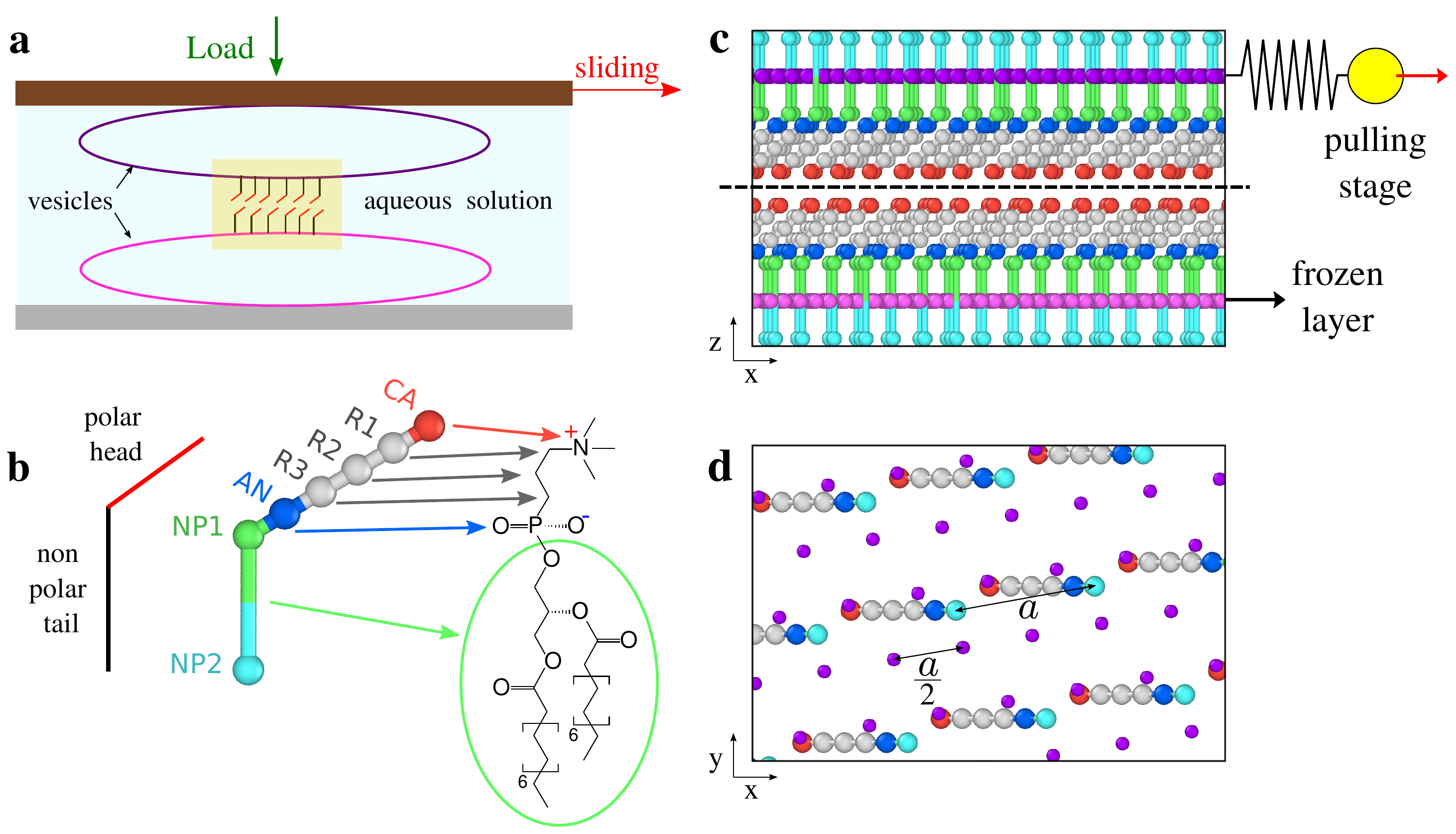}
\caption{\label{fig:system} (a) Sketch of the experimental SFA setup. 
(b) Three representations of one molecular unit in the vesicle wall.
Left: a radically simplified sticks scheme, as in panel (a).
Center: the simulated united-atom molecule. Right: the detailed chemical structure of
a
dipalmitoylphosphatidylcholine molecule. 
(c) The initial 
regularly spaced 
configuration of the system (side view).
Magenta, purple, red, blue, green and cyan spheres represent SUB and SUP layers, cations, anions, and NP1 and NP2 particles, respectively. Gray particles represent the R1, R2, and R3 neutral residues in between the cation and the anion. The yellow sphere represents the pulling stage advancing with constant speed. 
The dashed line marks the sliding interface. 
(d) Top view 
restricted to the particles above the black dashed line in (c); purple SUP particles are drawn smaller for better readability. 
}

\end{figure*}

\subsection{\label{sec:model}The Model}

We propose a model 
inspired by
surface force apparatus (SFA) experiments with confined 
self-assembled 
vesicles formed by organic polymers %consisting of
composed of hydrophobic tails and hydrophilic heads %composed by 
consisting of short zwitterionic chains \cite{Klein2016,Klein2019Langmuir,lin2020cells}.
The self-arranged vesicles can sit inside the SFA confined contact, as sketched in Figure~\ref{fig:system}a:
this arrangement provides a highly sensitive setup
for the measurement of the frictional shear stress between 
well-characterized flat
surfaces with molecular layers sticking out of 
them in mutual shearing motion and under
a controlled normal load \cite{lin-klein2021Review}.
The solid surfaces and the viscoelastic deformability of the vesicles 
cooperate in generating an essentially atomically flat interface 
between the exposed surfaces of two contacting vesicles.
This flat interface extends over the size of a vesicle, namely, several micrometers across.
All shearing occurs in this sliding interface, which is therefore responsible for all 
observed 
frictional forces \cite{goldberg2011,Klein2012polymers,Klein2019Langmuir,Klein2016,lin2020cells}.

A minimal model for simulating the frictional properties of this interface must include at least the exposed %polar heads 
zwitterionic head groups
of the 
molecules
%polymers constitutive of 
sticking out from
the vesicle, as sketched in Figure~\ref{fig:system}b.
We simulate these heads in a united-atom style as a string of 5 point-like particles.
From the interface point of view, the rest of the molecules, namely, the glycerol group and the long alkyl tails have the role of providing a directed support to the heads and transmitting the load and shear forces acted by the SFA setup.
In a coarse-grained representation of the hydrophobic inner part of the vesicle, we model this part of the system by parallel rigid layers, one for each contacting vesicle: we name them SUP and SUB layers.
%
%Both these layers are 
The lateral arrangement of molecules in these two layers is modeled using 
triangular lattices whose periodicity $a/2$ of $0.41$\,nm sets the equilibrium intermolecular spacing (see Figure~\ref{fig:system}d).
This spacing $a$ is the %typical 
characteristic
distance between neighboring molecules, matched to the typical areal number density of these vesicle-forming polymers, namely, 
$1.72$\,molecules/nm$^2$ \cite{tieleman1999}.
To prevent trivial and %artificial 
unrealistic 
perfect-commensuration effects, we impose a relative angle of rotation $\phi$ between 
the layers' crystalline directions (opposite rotations by $\pm\phi/2$ for each layer).
The value of $\phi=19.65^\circ$
and the numbers of lattice repetitions in the rigid layers are adapted in order to fit a supercell periodic in the horizontal $xy$ plane accommodating both lattices, as detailed in the 
Supporting Information.
We represent the experimental mesoscopic interface within a rectangular simulation supercell with %sides
dimensions
$l_x = 8.32$~nm and $l_y = 14.41$~nm.
The supercell contains $206$ molecules in each layer and 4 times as many atoms in each of the SUP and SUB rigid layers.

The link between the zwitterionic molecular part and the rigid layers is provided by the NP1 and NP2 units of each molecule that represent the nonpolar tails as a pair of point-like particles.
Each of these pairs of bonded atoms remains ``planted'' in one (out of four) of the %triangular spaces 
lattice nodes 
in either the SUP or the SUB layer.
The %layer
intralayer-molecule interaction 
results in keeping these nonpolar sections close to vertical alignment and regularly spaced, while allowing for a limited degree of elastic deformability;
see Figure~\ref{fig:system}b, c.

The SUB-layer particles (magenta particles in Figure~\ref{fig:system}c) are kept fully frozen.
Those in the SUP layer (purple particles in Figure~\ref{fig:system}c, d) are constrained to form an
identical, yet misaligned, 
rigid layer, allowed to translate in the three directions. 

Each embedded macromolecule %is 
therefore consists of   
a chain of seven particles, as depicted in Figure~\ref{fig:system}b. 
It %consists of 
starts with 
a cation (CA, red particle) followed by three uncharged residues (R1-R3, gray), an anion (AN, blue), and two 
uncharged particles, NP1 (green) and NP2 (cyan).
%With the example of 
Inspired by 
the dipalmitoylphosphatidylcholine molecule, %in mind
we set the masses (in a.m.u.) of the molecular beads
to 60 for CA, 15 for R1--R3, 80 for AN, and 50 for NP1 and NP2.
The CA and AN beads carry an associated charge of $q=0.25$ and $-q$, respectively, in 
elementary-charge 
units, while all other beads are neutral.
For 
comparison, we also consider a %fully %uncharged 
charge-free
version of the model with $q=0$ for all beads.
Successive atoms in %a 
each 
chain are connected by elastic springs %for 
representing 
both the stretching and the angular degrees of freedom.
All equilibrium angles $\theta_{\rm eq}$ are $180^\circ$, except for the NP2-NP1-AN angle, which we set to $111^\circ$, representative of a sp$^3$ skeleton oxygen, attempting to keep the %polar 
zwitterionic
heads tilted away from being vertical~\cite{Bockmann2008kinetics}.

The intramolecular harmonic interactions follow the standard expression
\begin{equation}\nonumber
	U_{\rm bond}\left( r \right) =\frac{1}{2}k_{\rm bond}\left(  r - r_{eq}\right)^{2}
\end{equation}
for linear springs, and
\begin{equation}\nonumber
    U_{\rm angle}\left( \theta \right) =\frac{1}{2}k_{\rm angle}\left(  \theta - \theta _{eq}\right)^{2}
\end{equation}
for angular springs. % Rob: we should check that our simulations don't suffer this problem: https://lammps.sandia.gov/threads/msg86464.html
Table~\ref{table:intra} lists
the parameters adopted for
both kinds
of intramolecular bonding interaction.

\begin{table}[t]
\caption{Intramolecular Interaction Parameters for the Bonded Interactions. All interactions not listed here involve non-bonded atoms and are of the Morse-type, Eq.~\eqref{eq:morse_pot}.}
\label{table:intra}
        \centering
        \renewcommand{\arraystretch}{1.3}  %vertical spacing in table
        \setlength{\tabcolsep}{4pt}        %horizontal spacing in table
        \begin{tabular}{|c|c|c|c|}
        \hline
        \multicolumn{4}{|c|}{harmonic bonds}\\
        \hline
        particle 1 & particle 2 &
        \begin{tabular}{@{}c@{}}$k_{\rm bond}$ \\ (N$\cdot$\,m$^{-1}$) \end{tabular} &
        \begin{tabular}{@{}c@{}}$r_{\rm eq}$ \\ (nm) \end{tabular} \\ \hline
        CA & R1 & \multirow{6}{*}{480} & \multirow{5}{*}{0.16} \\ \cline{1-2}
        R1 & R2 &  &  \\ \cline{1-2}
        R2 & R3 &  &  \\ \cline{1-2}
        R3 & AN &  &  \\ \cline{1-2}
        AN & NP1 & &  \\ \cline{1-2}\cline{4-4}
        NP1& NP2 & & 0.67 \\ \hline
        \end{tabular}
        %\quad
        \vspace{2mm}
    
        \begin{tabular}{|c|c|c|c|c|}
        \hline
        \multicolumn{5}{|c|}{harmonic angular interactions}\\
        \hline
         \begin{tabular}{@{}c@{}}particle  1\end{tabular} &
         \begin{tabular}{@{}c@{}}particle  2\end{tabular} &
         \begin{tabular}{@{}c@{}}particle  3\end{tabular} &
            \begin{tabular}{@{}c@{}}$k_{\rm angle}$ \\ (eV$\cdot$rad$^{-2}$) \end{tabular} &
            \begin{tabular}{@{}c@{}} $\theta_{\rm eq}$ \\ (degree)\end{tabular} \\ \hline
        CA & R1 & R2 & \multirow{4}{*}{20} & \multirow{4}{*}{180} \\ \cline{1-3}
        R1 & R2 & R3 &  &  \\ \cline{1-3}
        R2 & R3 & AN &  &  \\ \cline{1-3}
        R3 & AN & NP1 &  & \\ \cline{1-5} 
        AN & NP1 & NP2 & 2 & 111 \\ \hline
        \end{tabular}
\end{table}

For the non-bonded pairwise particle-particle interactions 
we adopt a Morse potential 
\begin{equation}\nonumber
    V_\text{Morse}\left(r \right) 
    = D_{0}\left[  e^{-2\alpha\left(r-r_{0} \right)}-2e^{-\alpha\left( r-r_{0}\right)}\right]
    \label{eq:morse_pot}
    %V\left(r \right) 
    %&= \phi\left(r \right)-\phi\left(R_c \right)-\left( r-R_c \right) \left. \frac{d\phi}{dr} \right|_{r=R_c} r<R_c
\end{equation}
with a standard shift and a linear term added so that both potential energy and force drop to zero at a cutoff distance $R_c$, see SI.
The stiffness parameter
$\alpha= 15$\,nm$^{-1}$ and the cutoff distance $R_c=1$\,nm are the same for all interaction pairs.
The values of
the potential well depth
$D_0$ 
and 
equilibrium spacing
$r_0$ are listed in Table~\ref{table:interactions}.
Note that, 
while all non-bonded beads interact pairwise through Morse terms,
as an exception, cross-layer interactions are restricted to the polar heads of the molecules (CA, R1-R3 and AN spheres in Figure~\ref{fig:system}), with spurious cross-layer terms removed by the $D_0=0$ values in Table~\ref{table:interactions}.

\begin{table}[t]
\centering
\renewcommand{\arraystretch}{1.3}  %vertical spacing in table
\setlength{\tabcolsep}{4pt}        %horizontal spacing in table
\begin{tabular}{|c|c|c|c|}
\hline
particle 1  & particle 2 & 
\begin{tabular}{@{}c@{}} $D_0$ \\ (eV)\end{tabular} & 
\begin{tabular}{@{}c@{}} $r_0$ \\ (nm)\end{tabular}  \\ \hline
\multicolumn{2}{|c|}{default} & 0.010 & 0.41   \\ \hline
SUP    & NP1/2 (SUP layer)   & 5.0 & \multirow{5}{*}{0.41} \\ \cline{1-3}
SUP    & NP1/2 (SUB layer)   & 0.0 &  \\ \cline{1-3}
SUB    & NP1/2 (SUB layer)   & 5.0 & \\ \cline{1-3}
SUB    & NP1/2 (SUP layer)   & 0.0 &  \\ \cline{1-3}
SUB    & SUP & 0.0 &  \\\hline
NP1 (SUP layer)   & NP1 (SUP layer)  &    5.0  & \multirow{6}{*}{0.82} \\ \cline{1-3}
NP1 (SUP layer)   & NP1 (SUB layer)  &    0.0  & \\ \cline{1-3}
NP1 (SUB layer)   & NP1 (SUB layer)  &    5.0  & \\ \cline{1-3}
NP2 (SUP layer)   & NP2 (SUP layer)  &    5.0  & \\ \cline{1-3}
NP2 (SUP layer)   & NP2 (SUB layer)  &    0.0  & \\ \cline{1-3}
NP2 (SUB layer)   & NP2 (SUB layer)  &    5.0  & \\ \hline
\end{tabular}
\caption{Values for the Morse-interactions parameters between pairs of particles.}
\label{table:interactions}
\end{table}

Within the cutoff radius,
Coulombic pairwise interactions are computed directly 
in real space,
while outside that distance
interactions are evaluated in reciprocal space.
For the reciprocal space, a particle-particle particle-mesh solver (PPPM)
\cite{hockney1989pppm,pollocl1996pppm}
is used with a precision of $10^{-4}$\,eV$\cdot$\,nm$^{-1}$,
which proved to be sufficiently accurate; see further details in the Supporting Information.

\subsection{\label{sec:simulations}Simulations}

We adopt LAMMPS \cite{lammps}
as the simulation platform for
integrating the equations of motion
\begin{align}
    m_i \ddot{r}_{i u} = F_{i u} -\gamma_{i u} m_i \dot{r}_{i u} +\xi_{i u}\,.
\end{align}
In addition to the conservative forces $F_{i u}$ explicitly provided by the force fields described in the previous section, 
%we need to model interactions with a solvent which we keep implicit, and do not include in the simulation. The solvent is simulated by
we impose a finite temperature using 
a Langevin thermostat with a damping rate $\gamma_{i u}$ 
applied to all particles forming the molecules and Gaussian random forces $\xi_{i u}$ 
\cite{Allen91}.
This thermostat is set to act only along the coordinates $ u =y,z$ in order to prevent any 
spurious thermostat-originated frictional 
damping along the most relevant sliding direction $ u =x$ \cite{Robbins01,rottler2003growth}.
Figure~S1 illustrates the robustness of the friction simulated in our model against the precise value of $\gamma_{i y} = \gamma_{i z} = \gamma $ adopted.
Eventually, we select a value of $\gamma=1$\,ps$^{-1}$ for all simulations.

While experiments with this kind of setup are usually carried out in (typical aqueous) solution, here, with the aim of providing a qualitative phenomenology (independent of the specific solvent nature), we adopt a suitably enlarged residue size $r_0$ and exploit a Langevin approach effectively taking care of degrees of freedom inherent in the real, physical system, which are not explicitly included in our model.\cite{galuschko2010}
Besides, a real solvent introduces electrostatic screening, which is partly accounted for in this model by the relatively small charges on the AN and CA residues.
As we intend to address a general mechanism without focusing on a specific system, we leave out all distance, frequency, and temperature dependence of the screening that a real solvent would entail.

For each simulation, we prepare an initial configuration by executing a sufficiently long ``running in'' simulation starting from the initial state shown in Figure~\ref{fig:system}c, letting the dynamics evolve with the appropriate load, temperature, and
fixed sliding velocity of the SUP layer 
until a steady state is reached.
In the appropriate
steady-state configuration, we attach a 
pulling stage  
to the SUP layer (yellow sphere in Figure~\ref{fig:system}c) through a spring of stiffness $k = 1$\,eV$\cdot$nm$^{-2} \simeq  0.16$\,N$\cdot$m$^{-1}$,
equivalent to a shear stress per unit elongation $k/\text{(supercell area)}=1.34 \times 10^9$\,MPa$\cdot$m$^{-1}$. 
We carry out the simulations with the 
stage 
advancing at constant speed in the $x$ direction: $x_{\rm stage} =  v_{\rm stage}\, t$. 
A default 
$v_{\rm stage} = 5$\,m$\cdot$s$^{-1}$, 
in the range of a typical MD approach,
is adopted, 
but we explore other velocities too.
In each simulation, the total advancement of the stage amounts to $100$\,nm.
We obtain the instantaneous shear stress from the spring elongation.
We start averaging this shear stress when the system enters a steady sliding state until the end of the simulation. This corresponds to one discarding an initial transient of $20$--$25$\,nm until at least the first slip event takes place. 

We report the averaged shear stress with vertical
bars reflecting the root mean squared fluctuations observed along the corresponding friction trace.
Large bars indicate stick-slip dynamics, while small bars originate from smooth sliding.

We apply relatively moderate values of loads ($L$) in the $0$--$20$\,MPa range, relevant for SFA experiments 
on organic %brushes
macromolecules.

\section{\label{sec:results}Results}

\begin{figure*}[tb!]
\includegraphics[width=0.8\textwidth]{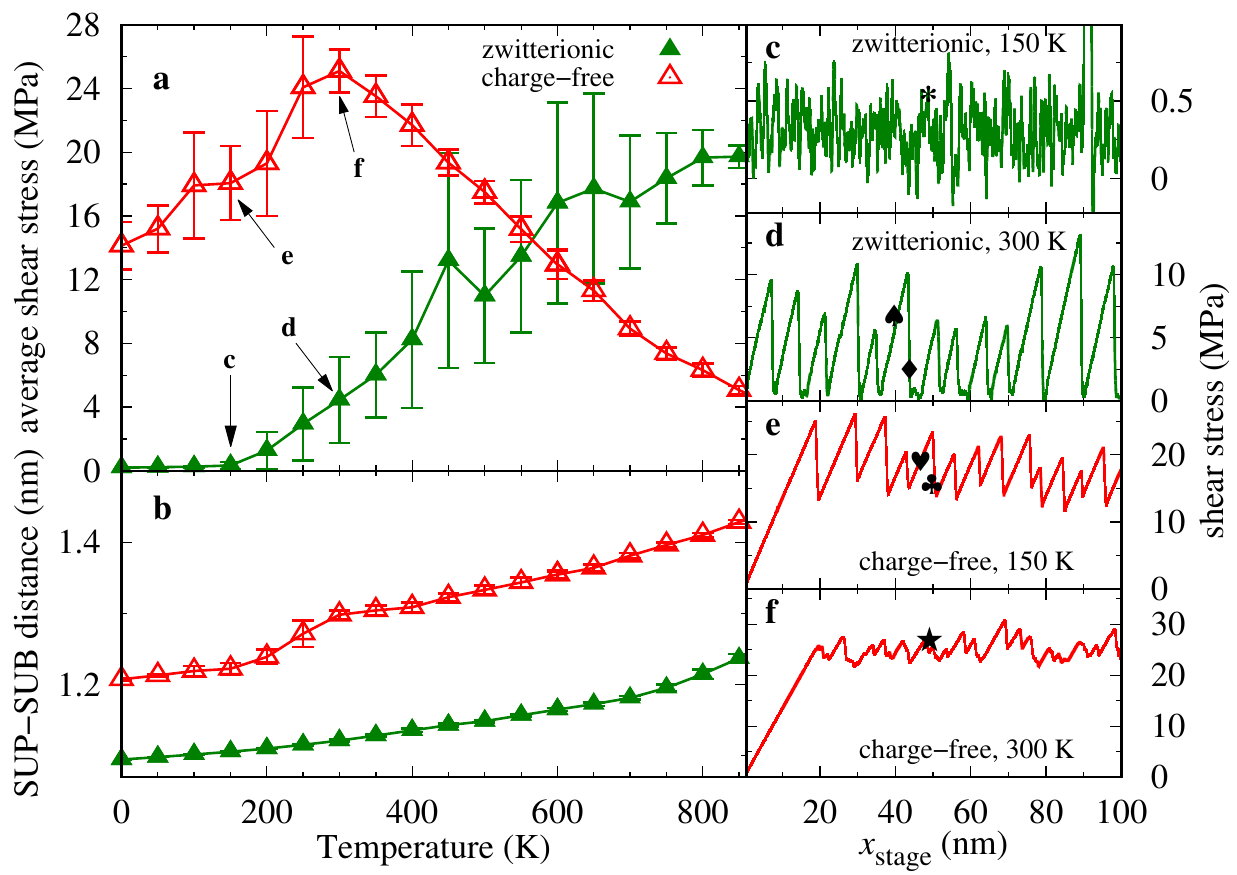}
\caption{
    \label{fig:cycles_T}
    (a) Nonmonotonic 
    variation of the 
    frictional shear stress and 
    (b) the distance between the rigid layers as a function of temperature for %charged 
    zwitterionic
    and %uncharged 
    charge-free
    systems. 
    $v_{\rm stage} = 5$\,m$\cdot$s$^{-1}$, $L = 10$\,MPa. 
    (c-f) Shear traces for the pointed temperatures in panel (a). 
    Symbols in (c-f) refer to the snapshots in Figure~\ref{fig:snaps}.
}
\end{figure*}

\begin{figure*}[tb!]
\includegraphics[width=\textwidth]{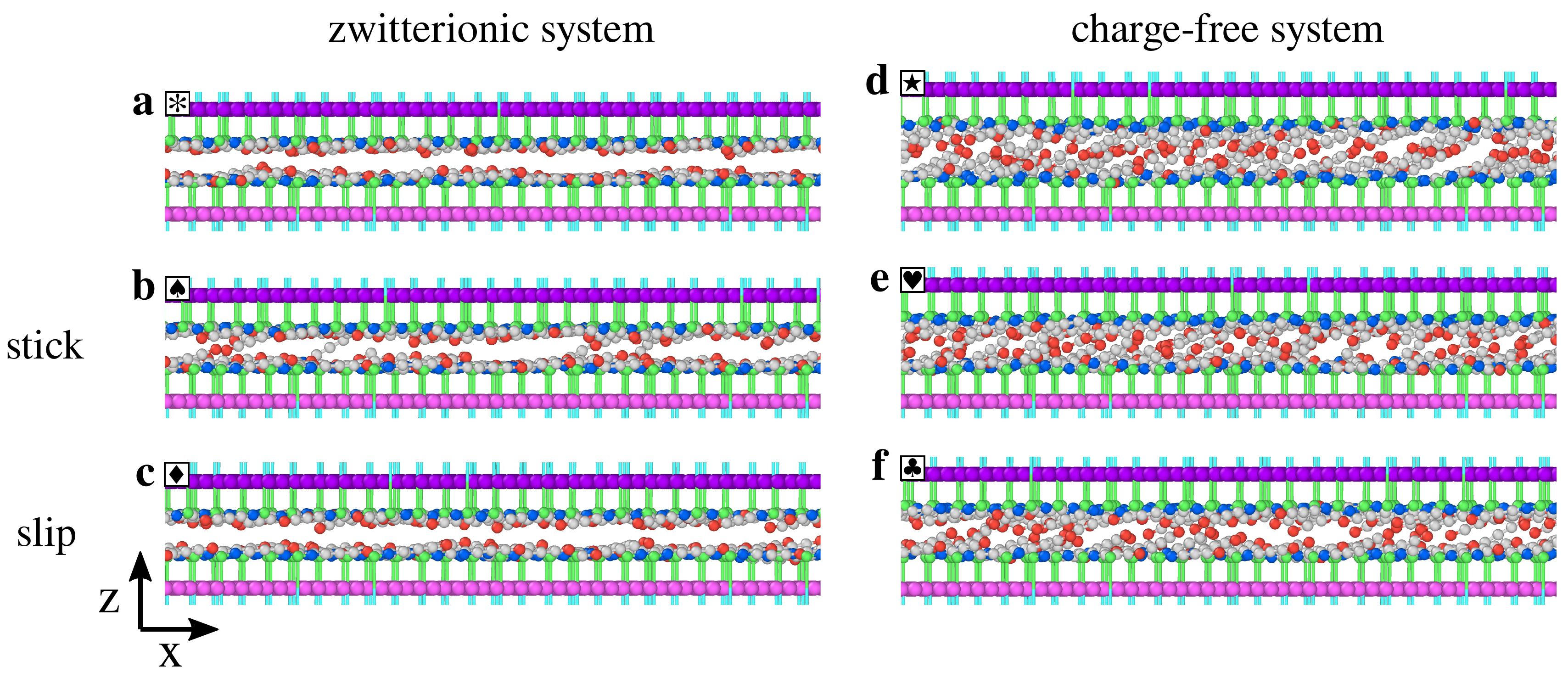}
\caption{
    \label{fig:snaps} Side views of 5~nm $y$-thick slices of simulation snapshots. 
    Each snapshot corresponds to the time instant marked by the corresponding symbol in panels c-f of Figure~\ref{fig:cycles_T}. 
    (a) Smooth sliding, %charged 
    zwitterionic
    system, $T = 150$\,K, (b) stick point, %charged 
    zwitterionic
    system, $T = 300$\,K, (c) slip point, %charged 
    zwitterionic
    system, $T = 300$\,K, (d) high friction state, %uncharged 
    charge-free
    system, $T = 300$\,K, (e) stick point, %uncharged 
    charge-free
    system, $T = 150$\,K, and (f) slip point, %
    charge-free
    %uncharged 
    system, $T = 150$\,K.
}
\end{figure*}

\begin{figure*}[tb!]
\includegraphics[width=\textwidth]{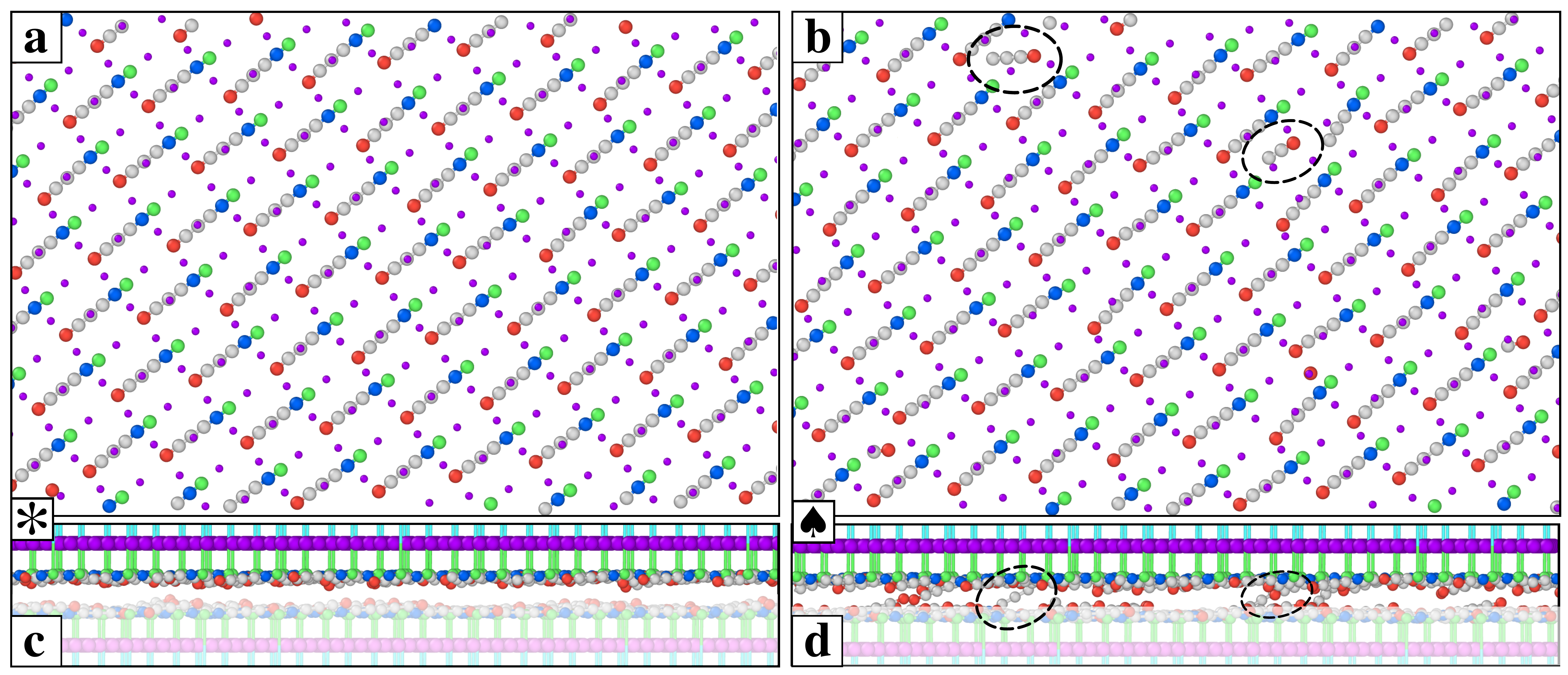}
\caption{
    \label{fig:tops} 
    (a, b) Top views of the snapshots of Figure~\ref{fig:snaps}a ($150$\,K) and \ref{fig:snaps}b ($300$\,K), 
    including a horizontal slice from the sliding plane in between the chains to immediately above the rigid SUP layer. This slice is the unshaded part of the side views (c) and (d). Black dashed circles highlight SUB chains intersecting the SUP chains' cation plane. 
    The end cation of each of those SUB chains, feels a relatively strong Coulomb interaction with the two top-chain anions, generating the interlocking spots responsible for the stick. }
\end{figure*}

\begin{figure*}
\includegraphics[width=\columnwidth]{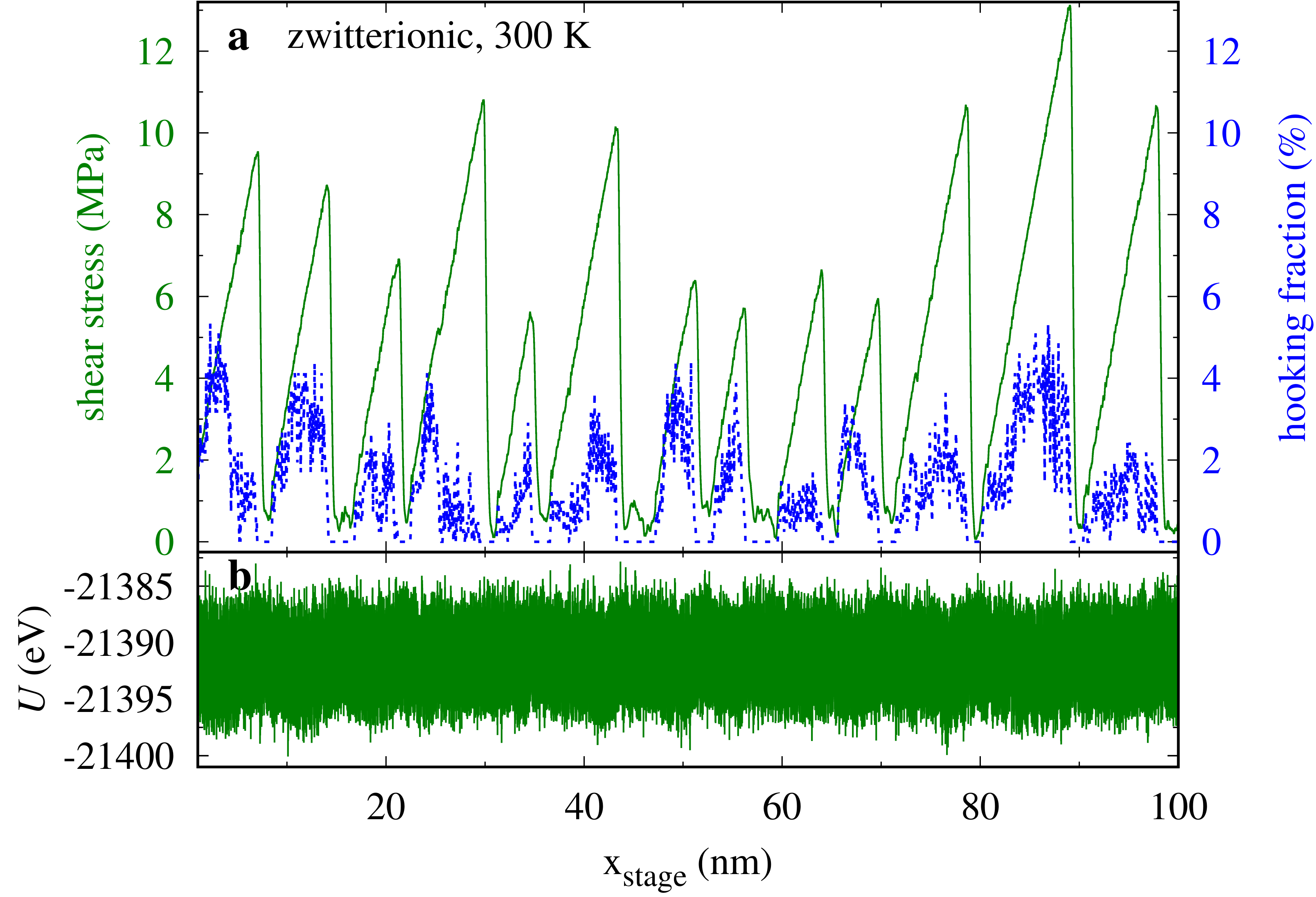}
\caption{
    \label{shear-h_charged} (a) Percentile hooking fraction $h$ as a function of the stage displacement correlated with the frictional shear stress for the same simulation as in Figure~\ref{fig:cycles_T}d. (b) The total potential energy $U$ for the same simulation.}
\end{figure*}

\begin{figure*}[tb!]
\includegraphics[width=0.8\textwidth]{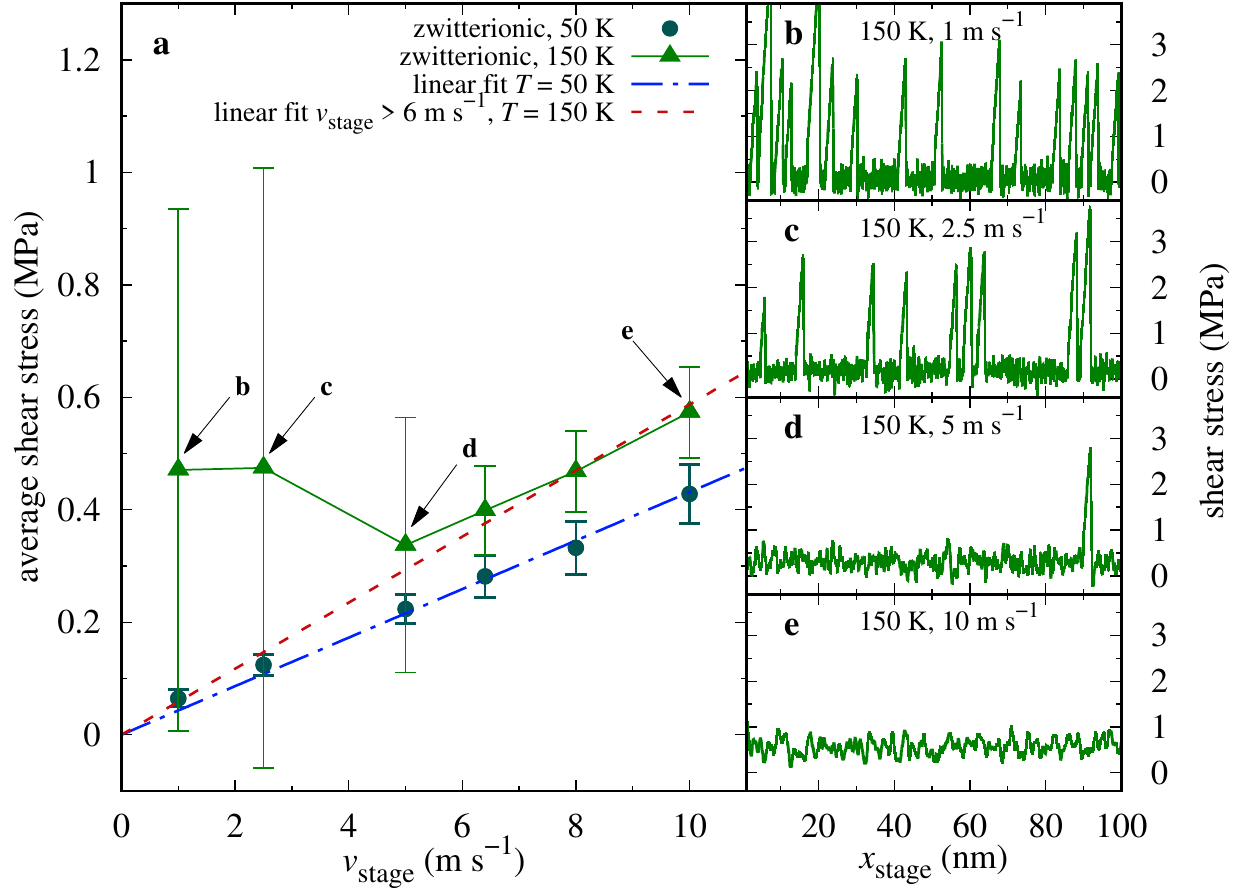}
\caption{
    \label{fig:vX_0.25} Stick-slip to smooth sliding transition as a function of velocity. 
    (a) The frictional shear stress as a function of the sliding velocity, %charged 
    zwitterionic
    system, $L = 10$\,MPa.
    (b-e) Shear traces for the pointed velocities in panel (a) at 150\,K. 
    Dashed line: fit of $T = 150$\,K and $v_{\rm stage} > 6$\,m$\cdot$s$^{-1}$. 
    Dot-dashed line: fit of the $T = 50$\,K, including all points. 
}
\end{figure*}

Figure~\ref{fig:cycles_T}a displays the frictional shear stress as a function of temperature.
Consider first the green symbols, reporting the simulations of the default model, the one involving %nonzero charges
zwitterionic head groups. 
At low temperature, a smooth-sliding regime (Figure~\ref{fig:cycles_T}c) characterized by extremely small friction is observed. 
In the smooth-sliding regime, the two layers remain substantially flat and well ordered 
due to the Coulombic interactions between cations and anions in the same layer 
(see Figures~\ref{fig:snaps}a and \ref{fig:tops}a):
chains of opposite layers do not entangle, and they slide on top of each other encountering a quite small corrugation due to the discommensuration associated with the mutual angular misalignment.
Starting from approximately $T \geq 200$\,K, stick-slip dynamics sets in (see Figure~\ref{fig:cycles_T}d), and friction increases substantially.
As temperature is raised, thermal fluctuations promote out-of-plane chain movements leading to transient interlocking (see Figures~\ref{fig:snaps}b and \ref{fig:tops}b).
%These 
The cationic 
chain head reaching 
through the opposite layer forms 
transient bonds
with the 
anions belonging to two 
adjacent chains in the countersurface. 
The fraction 
of these bonds can be quantified through the ``hooking fraction'' $h$, i.e., the degree of interpenetration, 
defined quantitatively in the Supporting Information and reported in Figure~\ref{shear-h_charged}a 
for 
the $300$\,K dynamics of Figure~\ref{fig:cycles_T}d.
These bonds are responsible for the ``stick'' intervals, where the SUP remains essentially static and the driving spring elongates. 
The shear stress drops to nearly 0 after each slip and so does the hooking fraction (see Figure~\ref{fig:snaps}c and Figure~\ref{shear-h_charged}a). 
The smooth-sliding and stick-slip motions of Figure~\ref{fig:cycles_T}c,d can be inspected in short movies reporting the last 6\,ns of the simulations 
(i.e.\ the last 30\,nm of the stage advancement in the shear traces), available as Movies S1 and S2.
The transition from  
smooth sliding to stick-slip that we 
observe for increasing temperature 
depends on the sliding velocity and on the stiffness of the driving spring with smaller velocities and softer springs favoring stick-slip over smooth sliding \cite{VanossiRMP13,Dong11}.
An important parameter of a tribological contact is 
the critical velocity $v_c$ above which intermittent stick-slip dynamics tends to disappear.
In our simulations, we can observe this 
disappearance as a function of $v_\text{stage}$ at $T = 150$\,K in Figure~\ref{fig:vX_0.25}. 
With the adopted model parameters, 
our simulations show clear stick-slip dynamics for $v_\text{stage}< 3$\,m$\cdot$s$^{-1}$ and smooth sliding for $v_\text{stage}> 6$\,m$\cdot$s$^{-1}$.
The model therefore predicts a critical velocity of $\simeq 5$\,m$\cdot$s$^{-1}$. 
However, $v_c$ can change by several orders of magnitude 
depending on system parameters. 
For example, at $T = 50$\,K (see dot-dashed line in Figure~\ref{fig:vX_0.25}a), the critical velocity is extremely small, and simulations capable of observing stick-slip would be far too long.
It is therefore unfeasible to evaluate $v_c$ systematically through simulations. 
Experimentally, in
ref~\onlinecite{Drummond01}, this critical velocity is reported for SFA experiments involving squalane films. 
In those experiments, an increase of $v_c$ with increasing temperature was observed: that result is compatible with the outcome of the present model.

Coming back to Figure~\ref{fig:cycles_T}a, as temperature increases to $T \simeq 600$\,K, friction increases less and less until it peaks near $800$~K. 
The transient bonds are numerous and relatively short-lived. 
The energy of each one of such bonds can be estimated (neglecting the small Morse contributions) by the difference in Coulomb attraction of a cation placed in between two adjacent 
anions of the opposite layer (distance $\simeq 0.41$~nm) and placed in its flat-layer configuration (distance $\simeq 0.51$~nm), 
which gives $\simeq 85$~meV. 
This transient bond energy is close to the thermal energy $k_\text{B}T\simeq86$\,meV for $T \simeq 1000$\,K, precisely in the temperature region of the observed friction peak.
For even higher $T > 1000$~K, these bonds are destabilized and eventually friction decreases. 
In order to analyze the effect of the molecular charges on friction, we run analogous simulations for the %uncharged 
charge-free
system, reported as red curves in Figure~\ref{fig:cycles_T}. 
Remarkably, across the temperature range from $T = 0$\,K to $T = 550$\,K, friction is significantly larger than that for the %charged 
zwitterionic
system.
The reason for this difference is that, even at $0$\,K, both molecular layers are significantly disordered due to the lack of long-range interactions.
The chain-orientation disorder leads to a tilt-angle disorder too and to a significant corrugation of the two mutually sliding layers of the molecular heads (CA). 
This corrugation is also reflected in the consistently larger average SUP-SUB distance, shown in Figure~\ref{fig:cycles_T}b, compared to that in the %charged 
zwitterionic
case.
Due to this extra corrugation, the hooked fraction of the %uncharged 
charge-free
system remains significant, even down to low temperatures. 

The shear traces for the %uncharged 
charge-free system 
(Figure~\ref{fig:cycles_T}e,f) exhibit stick-slip of smaller amplitude than those for the %charged 
zwitterionic system. 
As illustrated in Figure~\ref{fig:snaps}d,e and in Movies S1, S2, S3 and S4, this is due to the %uncharged brushes  
charge-free molecules 
showing a larger density of protruding chains ready to interlock before the slip event has exhausted the energy stored in the pulling spring (Figure~\ref{fig:snaps}f).
For $T > 300$\,K, thermal fluctuations start to undermine the weaker Morse-type bonds of the protruding chains leading to a progressive decrease in friction. 

\begin{figure*}[tb!]
\includegraphics[width=0.8\textwidth]{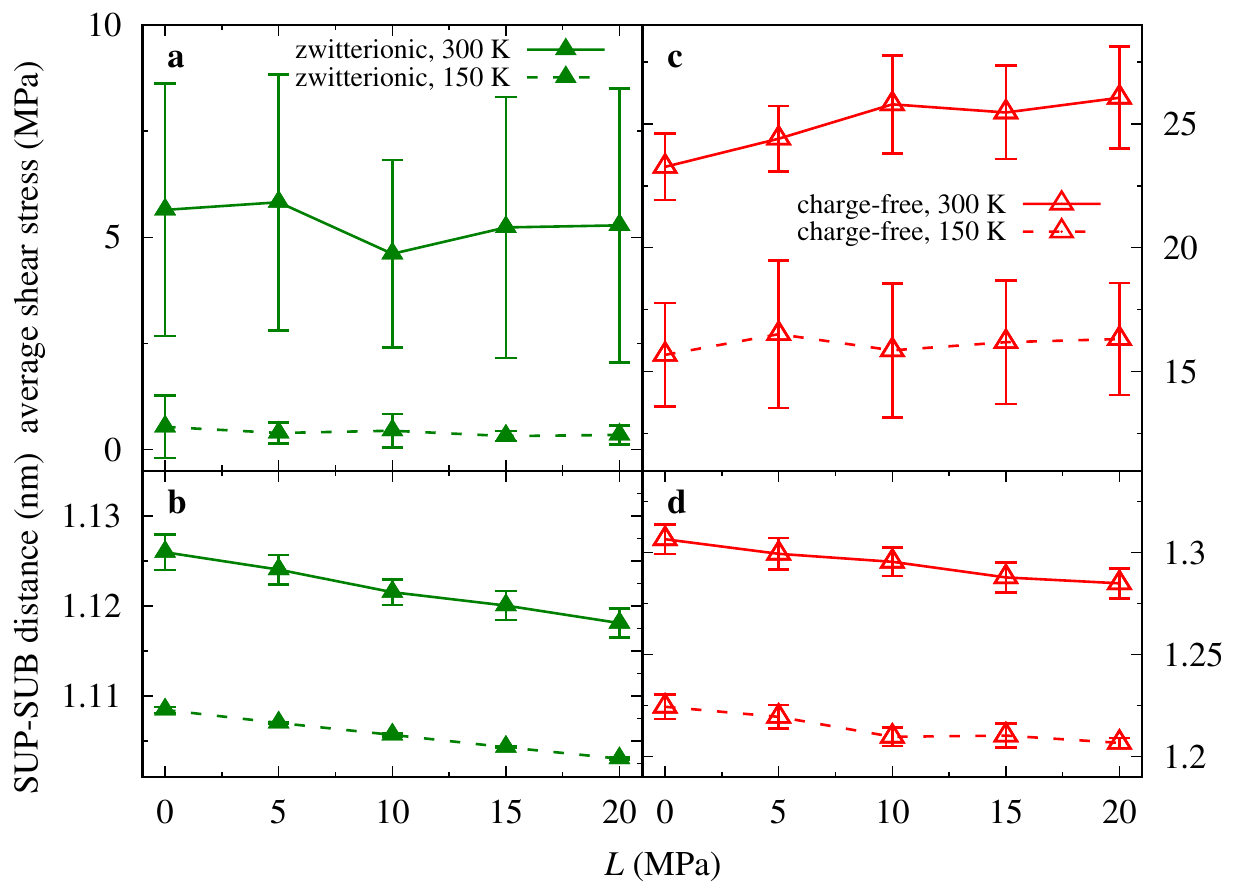}
\caption{
    \label{fig:cycles_load} (a,c) Frictional shear stress and (b,d) the average distance between the rigid layers as a function of load $L$, with $v_{\rm stage} = 5$\,m$\cdot$s$^{-1}$.
}
\end{figure*}

So far, the applied load 
$L$ 
was fixed to 10~MPa. 
To explore how $L$ affects the discussed phenomenology, we perform friction load cycles between 0 and 20~MPa and then back to 0 MPa as reported in Figure~\ref{fig:cycles_load}.
The outcome of these simulations indicates that the shear stress 
does not change significantly   
upon loading, despite the chain-layer compression visible in Figure~\ref{fig:cycles_load}b,d. 
Additionally, the unloading data retrace those of the loading simulations with no visible hysteresis.
For this reason, each point in Figure~\ref{fig:cycles_load} is obtained as an average over both loading and unloading traces for that given load.
The frictional shear traces for different loads are reported in Figures~S2 and S3. 
The traces of the %charged 
zwitterionic
system, Figure~S2, show that 
the critical velocity remains practically unchanged for all the investigated loads at $T\simeq150$\,K.
Given such a weak friction dependence on load, it is essentially meaningless to define a friction coefficient for this model.
From Figure~\ref{fig:cycles_load}, we expect the same conclusion at any $T$ in the considered range.
The %uncharged 
charge-free reference 
system exhibits a marginally significant friction increase with load.

As for the velocity dependence, in the stick-slip regime, it is also expected to be very mild. Indeed, Figure~S4 shows essentially no dependence as long as the dynamics is stick-slip. 
The %charged 
zwitterionic
system at $T=150$\,K reaches smooth sliding for $v_{\rm stage} > 6$\,m$\cdot$s$^{-1}$. 
The resulting friction linear increase, practically invisible in Figure~S4, is clear in Figure~\ref{fig:vX_0.25}a. 
Likewise, the smooth-sliding dynamics at 50\,K also produces velocity-linear friction over all simulated velocities.
In contrast, the %uncharged 
charge-free
system has stick-slip dynamics at all simulated temperatures resulting in velocity-independent friction.

Figure~\ref{shear-h_charged} and analogous plots for different dynamical conditions exhibit clear signs of 
correlation between the hooking fraction $h$ and the frictional shear stress.
We expect that $h$ should 
also correlate with the total potential energy $U$. 
Specifically, we expect %reduction 
a decrease in total potential energy 
as the number of 
hooked stick 
points 
increases. 
This anticorrelation is illustrated in Figure~S5 for the %uncharged
charge-free
system at $T=0$\,K.
However, at $T=300$\,K, the total potential energy is extremely noisy due to thermal fluctuations (see Figure~\ref{shear-h_charged}), and these anticorrelations 
are hard to detect visually.

\begin{figure*}[tb!]
\includegraphics[width=0.9\textwidth]{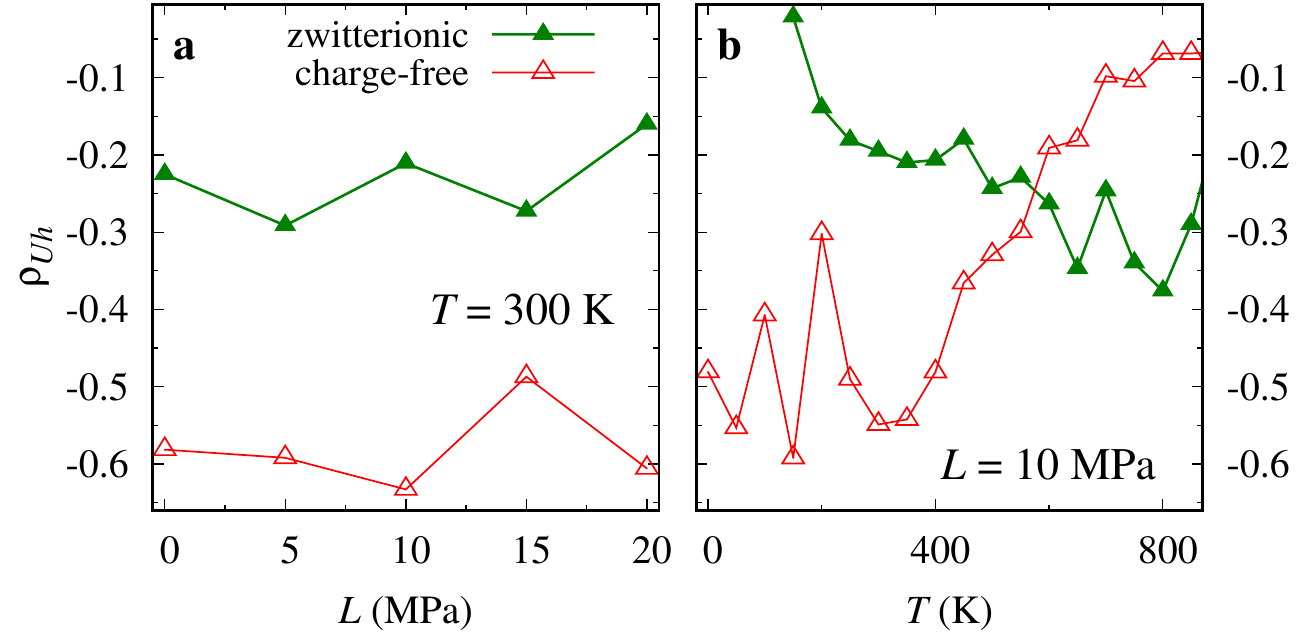}
\caption{
    \label{fig:corr-load-300} Correlation coefficient $\rho_{U h}$ between the hooked fraction and total potential energy, eq.~\ref{eq:rho}, for the %charged 
    zwitterionic
    and %uncharged 
    charge-free
    systems at $v_{\rm stage} = 5$\,m$\cdot$s$^{-1}$ as a function of (a) the applied load and (b) the temperature. As in smooth sliding $h \equiv 0$, $\rho_{U h}$ is defined for stick-slip dynamics only. }
\end{figure*}

Figure S6 illustrates these correlations with scatter plots for the %charged 
zwitterionic
(panel a) and %uncharged 
charge-free
(panel b) models at $T = 300$\,K, related to the traces of Figure~\ref{fig:cycles_T}d,f.
These scatter plots provide qualitative hints of these correlations. 
For a quantitative evaluation of these correlations, we calculate the Pearson correlation coefficient \cite{taylor1997introduction} 
\begin{equation}
	\rho_{U h}=\frac{\sum U_{t}h_{t}-i\bar{U}\bar{h}}{\sqrt{\left( \sum U^{2}_{t}-i \bar{U}^{2}\right) \left( \sum h^{2}_{t}-i \bar{h}^{2}\right)}},\label{eq:rho}
\end{equation}
and report it as a function of load and temperature in Figure~\ref{fig:corr-load-300}.
$\rho_{U h}$ is systematically negative, confirming the expected anticorrelation.
At $T=300$\,K the %uncharged 
charge-free
model exhibits more negative anticorrelation compared to the %charged 
zwitterionic
model at all the investigated loads (Figure~\ref{fig:corr-load-300}a). 
As a function of temperature, so far, the %charged 
zwitterionic
and %uncharged 
charge-free
models behave quite differently. 
The %charged 
zwitterionic
model has null $h$ at low temperatures (smooth sliding), and therefore, $\rho_{U h}$ is undefined.
As stick-slip develops, $\rho_{U h}$ becomes more and more negative.
In contrast, the %uncharged 
charge-free
system exhibits stick-slip down to a temperature of zero with the correspondingly largely negative $\rho_{U h}$.
As temperature is raised, these correlations approach zero and correspondingly friction decreases (Figure~\ref{fig:cycles_T}a).

\section{Discussion and Conclusions}

In this work, we have introduced and studied a model that offers a microscopic implementation for 
friction mediated by thermally activated formation and rupture of interfacial contacts, 
a subject which so far has been explored by means of phenomenologic theory \cite{Filippov04,Barel10b,Guerra2016,Blass2017,Liu12a,Li11b,Tian2017,Ouyang2019,Shao2017}. 
This kind of theory has been found to have wide applications in describing friction and wear in dry and lubricated contacts over a broad range of lengths and revealed the origin of new, 
unexpected phenomena such as non-Amonton's variation of friction force with normal load and nonmonotonic dependence of friction on sliding speed and temperature. 
The microscopic model proposed here advances understanding mechanisms underlying the phenomenological theory and offers new pathways for the rational control of frictional response.

The main outcome of this model consists of a remarkable increase in friction with temperature. 
Depending on the operating conditions, sliding can occur with stick-slip or smooth advancement. The high-friction stick state results from the interlocking of %polymer 
molecular 
chains promoted by thermal fluctuations.
In the proposed model, Coulombic interactions between the %polymeric brushes
zwitterionic macromolecules 
enhance the %inter-polymer 
intermolecular 
interactions, favoring flat and ordered layers and smooth sliding at low temperatures.
We can imagine different types of %inter-polymer 
intermacromolecule 
interactions, for example, hydrogen bonding or solvent-mediated couplings,
that could support a similar kind of low-temperature ordering.
Thermally activated interlocking is therefore also likely to account for thermally enhanced friction in more general contexts, as for example, in the experiments of ref~\onlinecite{Drummond01}.

%Charged brushes 
Zwitterionic macromolecules 
are promising systems for the control of friction by externally applied electric fields.
The change in the orientation of the molecular %dipoles 
zwitterionic heads 
due to the field may affect interlocking and, therefore, friction\cite{karuppiah2009}.
We hope that this model stimulates further experiments in this and 
related 
%other 
directions.

\begin{acknowledgement}

The authors acknowledge support from the grant PRIN2017 UTFROM of the Italian Ministry of University and Research. A.V. also acknowledges support by ERC Advanced Grant ULTRADISS, contract No. 8344023, and by the European Union’s Horizon 2020 research and innovation programme under grant agreement No. 899285. M.U. acknowledges the financial support of the Israel Science Foundation, Grant 1141/18. 
All the authors thank 
Carlos Drummond,
Di Jin, Jacob Klein, Erio Tosatti, and Yu Zhang 
for the useful discussions. 

The authors declare no competing financial interest.
\end{acknowledgement}

\begin{suppinfo}

The Supporting Information is available free of charge at \url{https://pubs.acs.org/doi/10.1021/acs.jpcc.1c09542}

Details of the chain arrangement, supercell geometry, and intermolecular interactions; the average shear stress as a
function of the damping parameter $\gamma$ (Figure S1); the definition of the hooking fraction $h$; the frictional shear traces of
the zwitterionic (Figure S2) and charge-free (Figure S3) systems for the load dependence of Figure~\ref{fig:cycles_load}a,c; shear stress
and SUP-SUB distance dependence on the sliding velocity for zwitterionic and charge-free systems (Figure S4);
correlation of the hooking fraction, shear stress, and total potential energy for the charge-free system at $T=0$~K
(Figure S5); scatter plots illustrating the anticorrelation between the total potential energy and the hooking fraction
that determine the $T=300$~K points in Figure~\ref{fig:corr-load-300}b for zwitterionic and charge-free systems (Figure S6); technical
details about 4 short supporting movies illustrating the MD simulations corresponding to the last 30 nm of the stage's
displacement in the shear traces of Figure~\ref{fig:cycles_T}c-f (PDF)

Last 6 ns of the MD simulation corresponding to the force trace shown in Figure~\ref{fig:cycles_T}c (MP4)

Last 6 ns of the MD simulation corresponding to the force trace shown in Figure~\ref{fig:cycles_T}d (MP4)

Last 6 ns of the MD simulation corresponding to the force trace shown in Figure~\ref{fig:cycles_T}e (MP4)

Last 6 ns of the MD simulation corresponding to the force trace shown in Figure~\ref{fig:cycles_T}f (MP4)
\end{suppinfo}

%\bibliography{biblioMMG.bib,biblio.bib}

\begin{mcitethebibliography}{58}
\providecommand*\natexlab[1]{#1}
\providecommand*\mciteSetBstSublistMode[1]{}
\providecommand*\mciteSetBstMaxWidthForm[2]{}
\providecommand*\mciteBstWouldAddEndPuncttrue
  {\def\EndOfBibitem{\unskip.}}
\providecommand*\mciteBstWouldAddEndPunctfalse
  {\let\EndOfBibitem\relax}
\providecommand*\mciteSetBstMidEndSepPunct[3]{}
\providecommand*\mciteSetBstSublistLabelBeginEnd[3]{}
\providecommand*\EndOfBibitem{}
\mciteSetBstSublistMode{f}
\mciteSetBstMaxWidthForm{subitem}{(\alph{mcitesubitemcount})}
\mciteSetBstSublistLabelBeginEnd
  {\mcitemaxwidthsubitemform\space}
  {\relax}
  {\relax}

\bibitem[Wu \latin{et~al.}(2015)Wu, Wei, Cai, and Zhou]{wu2015}
Wu,~Y.; Wei,~Q.; Cai,~M.; Zhou,~F. Interfacial friction control. \emph{Adv.
  Mater. Interfaces} \textbf{2015}, \emph{2}, 1400392\relax
\mciteBstWouldAddEndPuncttrue
\mciteSetBstMidEndSepPunct{\mcitedefaultmidpunct}
{\mcitedefaultendpunct}{\mcitedefaultseppunct}\relax
\EndOfBibitem
\bibitem[Manini \latin{et~al.}(2017)Manini, Mistura, Paolicelli, Tosatti, and
  Vanossi]{Manini17}
Manini,~N.; Mistura,~G.; Paolicelli,~G.; Tosatti,~E.; Vanossi,~A. Current
  trends in the physics of nanoscale friction. \emph{Adv. Phys. X}
  \textbf{2017}, \emph{2}, 569--590\relax
\mciteBstWouldAddEndPuncttrue
\mciteSetBstMidEndSepPunct{\mcitedefaultmidpunct}
{\mcitedefaultendpunct}{\mcitedefaultseppunct}\relax
\EndOfBibitem
\bibitem[Vanossi \latin{et~al.}(2018)Vanossi, Dietzel, Schirmeisen, Meyer,
  Pawlak, Glatzel, Kisiel, Kawai, and Manini]{Vanossi18}
Vanossi,~A.; Dietzel,~D.; Schirmeisen,~A.; Meyer,~E.; Pawlak,~R.; Glatzel,~T.;
  Kisiel,~M.; Kawai,~S.; Manini,~N. Recent highlights in nanoscale and
  mesoscale friction. \emph{Beilstein J. Nanotechnol.} \textbf{2018}, \emph{9},
  1995--2014\relax
\mciteBstWouldAddEndPuncttrue
\mciteSetBstMidEndSepPunct{\mcitedefaultmidpunct}
{\mcitedefaultendpunct}{\mcitedefaultseppunct}\relax
\EndOfBibitem
\bibitem[Krim(2019)]{Krim19}
Krim,~J. Controlling friction with external electric or magnetic fields: 25
  examples. \emph{Front. Mech. Eng.} \textbf{2019}, \emph{5}, 22\relax
\mciteBstWouldAddEndPuncttrue
\mciteSetBstMidEndSepPunct{\mcitedefaultmidpunct}
{\mcitedefaultendpunct}{\mcitedefaultseppunct}\relax
\EndOfBibitem
\bibitem[Sang \latin{et~al.}(2001)Sang, Dub\'e, and Grant]{Sang01}
Sang,~Y.; Dub\'e,~M.; Grant,~M. Thermal Effects on Atomic Friction. \emph{Phys.
  Rev. Lett.} \textbf{2001}, \emph{87}, 174301\relax
\mciteBstWouldAddEndPuncttrue
\mciteSetBstMidEndSepPunct{\mcitedefaultmidpunct}
{\mcitedefaultendpunct}{\mcitedefaultseppunct}\relax
\EndOfBibitem
\bibitem[Dudko \latin{et~al.}(2002)Dudko, Filippov, Klafter, and
  Urbakh]{Dudko02}
Dudko,~O.~M.; Filippov,~A.; Klafter,~J.; Urbakh,~M. Dynamic force spectroscopy:
  a Fokker Planck approach. \emph{Chem. Phys. Lett.} \textbf{2002}, \emph{352},
  499--504\relax
\mciteBstWouldAddEndPuncttrue
\mciteSetBstMidEndSepPunct{\mcitedefaultmidpunct}
{\mcitedefaultendpunct}{\mcitedefaultseppunct}\relax
\EndOfBibitem
\bibitem[Szlufarska \latin{et~al.}(2008)Szlufarska, Chandross, and
  Carpick]{Szlufarska08}
Szlufarska,~I.; Chandross,~M.; Carpick,~R. Recent advances in single-asperity
  nanotribology. \emph{J. Phys. D} \textbf{2008}, \emph{41}, 123001\relax
\mciteBstWouldAddEndPuncttrue
\mciteSetBstMidEndSepPunct{\mcitedefaultmidpunct}
{\mcitedefaultendpunct}{\mcitedefaultseppunct}\relax
\EndOfBibitem
\bibitem[Brukman \latin{et~al.}(2008)Brukman, Gao, Nemanich, and
  Harrison]{Brukman08}
Brukman,~M.; Gao,~G.; Nemanich,~R.; Harrison,~J. Temperature dependence of
  single-asperity diamond-diamond friction elucidated using AFM and MD
  simulations. \emph{J. Phys. Chem. C} \textbf{2008}, \emph{112},
  9358--9369\relax
\mciteBstWouldAddEndPuncttrue
\mciteSetBstMidEndSepPunct{\mcitedefaultmidpunct}
{\mcitedefaultendpunct}{\mcitedefaultseppunct}\relax
\EndOfBibitem
\bibitem[Steiner \latin{et~al.}(2009)Steiner, Roth, Gnecco, Baratoff, Maier,
  Glatzel, and Meyer]{Steiner09}
Steiner,~P.; Roth,~R.; Gnecco,~E.; Baratoff,~A.; Maier,~S.; Glatzel,~T.;
  Meyer,~E. Two-dimensional simulation of superlubricity on NaCl and highly
  oriented pyrolytic graphite. \emph{Phys. Rev. B} \textbf{2009}, \emph{79},
  045414\relax
\mciteBstWouldAddEndPuncttrue
\mciteSetBstMidEndSepPunct{\mcitedefaultmidpunct}
{\mcitedefaultendpunct}{\mcitedefaultseppunct}\relax
\EndOfBibitem
\bibitem[Schirmeisen \latin{et~al.}(2006)Schirmeisen, Jansen, Holscher, and
  Fuchs]{Schirmeisen06}
Schirmeisen,~A.; Jansen,~L.; Holscher,~H.; Fuchs,~H. Temperature dependence of
  point contact friction on silicon. \emph{Appl. Phys. Lett.} \textbf{2006},
  \emph{88}, 123108\relax
\mciteBstWouldAddEndPuncttrue
\mciteSetBstMidEndSepPunct{\mcitedefaultmidpunct}
{\mcitedefaultendpunct}{\mcitedefaultseppunct}\relax
\EndOfBibitem
\bibitem[Barel \latin{et~al.}(2010)Barel, Urbakh, Jansen, and
  Schirmeisen]{Barel10a}
Barel,~I.; Urbakh,~M.; Jansen,~L.; Schirmeisen,~A. Multibond dynamics of
  nanoscale friction: the role of temperature. \emph{Phys. Rev. Lett.}
  \textbf{2010}, \emph{104}, 066104\relax
\mciteBstWouldAddEndPuncttrue
\mciteSetBstMidEndSepPunct{\mcitedefaultmidpunct}
{\mcitedefaultendpunct}{\mcitedefaultseppunct}\relax
\EndOfBibitem
\bibitem[Barel \latin{et~al.}(2010)Barel, Urbakh, Jansen, and
  Schirmeisen]{Barel10b}
Barel,~I.; Urbakh,~M.; Jansen,~L.; Schirmeisen,~A. Temperature dependence of
  friction at the nanoscale: when the unexpected turns normal. \emph{Tribol.
  Lett.} \textbf{2010}, \emph{39}, 311--319\relax
\mciteBstWouldAddEndPuncttrue
\mciteSetBstMidEndSepPunct{\mcitedefaultmidpunct}
{\mcitedefaultendpunct}{\mcitedefaultseppunct}\relax
\EndOfBibitem
\bibitem[Sheng and Wen(2012)Sheng, and Wen]{sheng2012electrorheological}
Sheng,~P.; Wen,~W. Electrorheological fluids: mechanisms, dynamics, and
  microfluidics applications. \emph{Annu. Rev. Fluid Mech.} \textbf{2012},
  \emph{44}, 143--174\relax
\mciteBstWouldAddEndPuncttrue
\mciteSetBstMidEndSepPunct{\mcitedefaultmidpunct}
{\mcitedefaultendpunct}{\mcitedefaultseppunct}\relax
\EndOfBibitem
\bibitem[Zhu \latin{et~al.}(2020)Zhu, Zeng, Wang, Yan, and He]{zhu2020nano}
Zhu,~J.; Zeng,~Q.; Wang,~Y.; Yan,~C.; He,~W. Nano-crystallization-driven high
  temperature self-lubricating properties of magnetron-sputtered WS 2 coatings.
  \emph{Tribol. Lett.} \textbf{2020}, \emph{68}, 1--11\relax
\mciteBstWouldAddEndPuncttrue
\mciteSetBstMidEndSepPunct{\mcitedefaultmidpunct}
{\mcitedefaultendpunct}{\mcitedefaultseppunct}\relax
\EndOfBibitem
\bibitem[Singh and Singh(2016)Singh, and Singh]{Singh2016}
Singh,~A.~K.; Singh,~T.~N. {Stability of the rate, state and temperature
  dependent friction model and its applications}. \emph{Geophys. J. Int.}
  \textbf{2016}, \emph{205}, 636--647\relax
\mciteBstWouldAddEndPuncttrue
\mciteSetBstMidEndSepPunct{\mcitedefaultmidpunct}
{\mcitedefaultendpunct}{\mcitedefaultseppunct}\relax
\EndOfBibitem
\bibitem[Tian \latin{et~al.}(2018)Tian, Goldsby, and Carpick]{Tian2018}
Tian,~K.; Goldsby,~D.~L.; Carpick,~R.~W. Rate and state friction relation for
  nanoscale contacts: thermally activated Prandtl-Tomlinson Model with chemical
  aging. \emph{Phys. Rev. Lett.} \textbf{2018}, \emph{120}, 186101\relax
\mciteBstWouldAddEndPuncttrue
\mciteSetBstMidEndSepPunct{\mcitedefaultmidpunct}
{\mcitedefaultendpunct}{\mcitedefaultseppunct}\relax
\EndOfBibitem
\bibitem[Tshiprut \latin{et~al.}(2009)Tshiprut, Zelner, and Urbakh]{Tshiprut09}
Tshiprut,~Z.; Zelner,~S.; Urbakh,~M. Temperature-induced enhancement of
  nanoscale friction. \emph{Phys. Rev. Lett.} \textbf{2009}, \emph{102},
  136102\relax
\mciteBstWouldAddEndPuncttrue
\mciteSetBstMidEndSepPunct{\mcitedefaultmidpunct}
{\mcitedefaultendpunct}{\mcitedefaultseppunct}\relax
\EndOfBibitem
\bibitem[Vanossi \latin{et~al.}(2013)Vanossi, Manini, Urbakh, Zapperi, and
  Tosatti]{VanossiRMP13}
Vanossi,~A.; Manini,~N.; Urbakh,~M.; Zapperi,~S.; Tosatti,~E. Colloquium:
  Modeling friction: From nanoscale to mesoscale. \emph{Rev. Mod. Phys.}
  \textbf{2013}, \emph{85}, 529\relax
\mciteBstWouldAddEndPuncttrue
\mciteSetBstMidEndSepPunct{\mcitedefaultmidpunct}
{\mcitedefaultendpunct}{\mcitedefaultseppunct}\relax
\EndOfBibitem
\bibitem[Manini \latin{et~al.}(2015)Manini, Braun, and Vanossi]{Manini15}
Manini,~N.; Braun,~O.~M.; Vanossi,~A. In \emph{Fundamentals of Friction and
  Wear on the Nanoscale 2nd ed.}; Gnecco,~E., Meyer,~E., Eds.; Springer,
  Berlin, 2015; p 175\relax
\mciteBstWouldAddEndPuncttrue
\mciteSetBstMidEndSepPunct{\mcitedefaultmidpunct}
{\mcitedefaultendpunct}{\mcitedefaultseppunct}\relax
\EndOfBibitem
\bibitem[Manini \latin{et~al.}(2016)Manini, Braun, Tosatti, Guerra, and
  Vanossi]{Manini16}
Manini,~N.; Braun,~O.~M.; Tosatti,~E.; Guerra,~R.; Vanossi,~A. Friction and
  nonlinear dynamics. \emph{J. Phys.: Condens. Matter} \textbf{2016},
  \emph{28}, 293001\relax
\mciteBstWouldAddEndPuncttrue
\mciteSetBstMidEndSepPunct{\mcitedefaultmidpunct}
{\mcitedefaultendpunct}{\mcitedefaultseppunct}\relax
\EndOfBibitem
\bibitem[Perkin and Klein(2013)Perkin, and Klein]{perkin2013}
Perkin,~S.; Klein,~J. Soft matter under confinement. \emph{Soft Matter}
  \textbf{2013}, \emph{9}, 10438--10441\relax
\mciteBstWouldAddEndPuncttrue
\mciteSetBstMidEndSepPunct{\mcitedefaultmidpunct}
{\mcitedefaultendpunct}{\mcitedefaultseppunct}\relax
\EndOfBibitem
\bibitem[Ma \latin{et~al.}(2019)Ma, Zhang, Yu, and Zhou]{ma2019brushing}
Ma,~S.; Zhang,~X.; Yu,~B.; Zhou,~F. Brushing up functional materials. \emph{NPG
  Asia Mater.} \textbf{2019}, \emph{11}, 1--39\relax
\mciteBstWouldAddEndPuncttrue
\mciteSetBstMidEndSepPunct{\mcitedefaultmidpunct}
{\mcitedefaultendpunct}{\mcitedefaultseppunct}\relax
\EndOfBibitem
\bibitem[Myshkin and Kovalev(2009)Myshkin, and Kovalev]{myshkin2009adhesion}
Myshkin,~N.~K.; Kovalev,~A.~V. \emph{Polymer tribology}; World Scientific,
  2009; pp 3--37\relax
\mciteBstWouldAddEndPuncttrue
\mciteSetBstMidEndSepPunct{\mcitedefaultmidpunct}
{\mcitedefaultendpunct}{\mcitedefaultseppunct}\relax
\EndOfBibitem
\bibitem[Chen \latin{et~al.}(2009)Chen, Briscoe, Armes, and Klein]{chen2009}
Chen,~M.; Briscoe,~W.~H.; Armes,~S.~P.; Klein,~J. Lubrication at physiological
  pressures by polyzwitterionic brushes. \emph{Science} \textbf{2009},
  \emph{323}, 1698--1701\relax
\mciteBstWouldAddEndPuncttrue
\mciteSetBstMidEndSepPunct{\mcitedefaultmidpunct}
{\mcitedefaultendpunct}{\mcitedefaultseppunct}\relax
\EndOfBibitem
\bibitem[Kreer(2016)]{kreer2016polymer}
Kreer,~T. Polymer-brush lubrication: a review of recent theoretical advances.
  \emph{Soft Matter} \textbf{2016}, \emph{12}, 3479--3501\relax
\mciteBstWouldAddEndPuncttrue
\mciteSetBstMidEndSepPunct{\mcitedefaultmidpunct}
{\mcitedefaultendpunct}{\mcitedefaultseppunct}\relax
\EndOfBibitem
\bibitem[De~Beer and M{\"u}ser(2013)De~Beer, and M{\"u}ser]{debeer2013}
De~Beer,~S.; M{\"u}ser,~M.~H. Alternative dissipation mechanisms and the effect
  of the solvent in friction between polymer brushes on rough surfaces.
  \emph{Soft Matter} \textbf{2013}, \emph{9}, 7234--7241\relax
\mciteBstWouldAddEndPuncttrue
\mciteSetBstMidEndSepPunct{\mcitedefaultmidpunct}
{\mcitedefaultendpunct}{\mcitedefaultseppunct}\relax
\EndOfBibitem
\bibitem[De~Beer \latin{et~al.}(2014)De~Beer, Kutnyanszky, Sch{\"o}n, Vancso,
  and M{\"u}ser]{deBeer2014}
De~Beer,~S.; Kutnyanszky,~E.; Sch{\"o}n,~P.~M.; Vancso,~G.~J.; M{\"u}ser,~M.~H.
  Solvent-induced immiscibility of polymer brushes eliminates dissipation
  channels. \emph{Nat. Commun.} \textbf{2014}, \emph{5}, 1--6\relax
\mciteBstWouldAddEndPuncttrue
\mciteSetBstMidEndSepPunct{\mcitedefaultmidpunct}
{\mcitedefaultendpunct}{\mcitedefaultseppunct}\relax
\EndOfBibitem
\bibitem[Raviv \latin{et~al.}(2003)Raviv, Giasson, Kampf, Gohy, J{\'e}r{\^o}me,
  and Klein]{raviv2003}
Raviv,~U.; Giasson,~S.; Kampf,~N.; Gohy,~J.-F.; J{\'e}r{\^o}me,~R.; Klein,~J.
  Lubrication by charged polymers. \emph{Nature} \textbf{2003}, \emph{425},
  163--165\relax
\mciteBstWouldAddEndPuncttrue
\mciteSetBstMidEndSepPunct{\mcitedefaultmidpunct}
{\mcitedefaultendpunct}{\mcitedefaultseppunct}\relax
\EndOfBibitem
\bibitem[Yu \latin{et~al.}(2012)Yu, Banquy, Greene, Lowrey, and
  Israelachvili]{yu2012langmuir}
Yu,~J.; Banquy,~X.; Greene,~G.~W.; Lowrey,~D.~D.; Israelachvili,~J.~N. The
  boundary lubrication of chemically grafted and cross-linked hyaluronic acid
  in phosphate buffered saline and lipid solutions measured by the surface
  forces apparatus. \emph{Langmuir} \textbf{2012}, \emph{28}, 2244--2250\relax
\mciteBstWouldAddEndPuncttrue
\mciteSetBstMidEndSepPunct{\mcitedefaultmidpunct}
{\mcitedefaultendpunct}{\mcitedefaultseppunct}\relax
\EndOfBibitem
\bibitem[R{\o}n \latin{et~al.}(2014)R{\o}n, Javakhishvili, Patil, Jankova,
  Zappone, Hvilsted, and Lee]{ron2014}
R{\o}n,~T.; Javakhishvili,~I.; Patil,~N.~J.; Jankova,~K.; Zappone,~B.;
  Hvilsted,~S.; Lee,~S. Aqueous lubricating properties of charged (ABC) and
  neutral (ABA) triblock copolymer chains. \emph{Polymer} \textbf{2014},
  \emph{55}, 4873--4883\relax
\mciteBstWouldAddEndPuncttrue
\mciteSetBstMidEndSepPunct{\mcitedefaultmidpunct}
{\mcitedefaultendpunct}{\mcitedefaultseppunct}\relax
\EndOfBibitem
\bibitem[Gaisinskaya-Kipnis and Klein(2016)Gaisinskaya-Kipnis, and
  Klein]{Klein2016}
Gaisinskaya-Kipnis,~A.; Klein,~J. Normal and frictional interactions between
  liposome-bearing biomacromolecular bilayers. \emph{Biomacromolecules}
  \textbf{2016}, \emph{17}, 2591--2602\relax
\mciteBstWouldAddEndPuncttrue
\mciteSetBstMidEndSepPunct{\mcitedefaultmidpunct}
{\mcitedefaultendpunct}{\mcitedefaultseppunct}\relax
\EndOfBibitem
\bibitem[Angayarkanni \latin{et~al.}(2019)Angayarkanni, Kampf, and
  Klein]{Klein2019Langmuir}
Angayarkanni,~S.~A.; Kampf,~N.; Klein,~J. Surface interactions between boundary
  layers of poly (ethylene oxide)--liposome complexes: Lubrication, bridging,
  and selective ligation. \emph{Langmuir} \textbf{2019}, \emph{35},
  15469--15480\relax
\mciteBstWouldAddEndPuncttrue
\mciteSetBstMidEndSepPunct{\mcitedefaultmidpunct}
{\mcitedefaultendpunct}{\mcitedefaultseppunct}\relax
\EndOfBibitem
\bibitem[Lin \latin{et~al.}(2020)Lin, Liu, Kampf, and Klein]{lin2020cells}
Lin,~W.; Liu,~Z.; Kampf,~N.; Klein,~J. The role of hyaluronic acid in cartilage
  boundary lubrication. \emph{Cells} \textbf{2020}, \emph{9}, 1606\relax
\mciteBstWouldAddEndPuncttrue
\mciteSetBstMidEndSepPunct{\mcitedefaultmidpunct}
{\mcitedefaultendpunct}{\mcitedefaultseppunct}\relax
\EndOfBibitem
\bibitem[Lin and Klein(2021)Lin, and Klein]{lin-klein2021Review}
Lin,~W.; Klein,~J. Recent progress in cartilage lubrication. \emph{Adv. Mater.}
  \textbf{2021}, \emph{33}, 2005513\relax
\mciteBstWouldAddEndPuncttrue
\mciteSetBstMidEndSepPunct{\mcitedefaultmidpunct}
{\mcitedefaultendpunct}{\mcitedefaultseppunct}\relax
\EndOfBibitem
\bibitem[Goldberg \latin{et~al.}(2011)Goldberg, Schroeder, Barenholz, and
  Klein]{goldberg2011}
Goldberg,~R.; Schroeder,~A.; Barenholz,~Y.; Klein,~J. Interactions between
  adsorbed hydrogenated soy phosphatidylcholine (HSPC) vesicles at
  physiologically high pressures and salt concentrations. \emph{Biophys. J.}
  \textbf{2011}, \emph{100}, 2403--2411\relax
\mciteBstWouldAddEndPuncttrue
\mciteSetBstMidEndSepPunct{\mcitedefaultmidpunct}
{\mcitedefaultendpunct}{\mcitedefaultseppunct}\relax
\EndOfBibitem
\bibitem[Klein(2012)]{Klein2012polymers}
Klein,~J. Polymers in living systems: from biological lubrication to tissue
  engineering and biomedical devices. \emph{Polym. Advan. Technol.}
  \textbf{2012}, \emph{23}, 729--735\relax
\mciteBstWouldAddEndPuncttrue
\mciteSetBstMidEndSepPunct{\mcitedefaultmidpunct}
{\mcitedefaultendpunct}{\mcitedefaultseppunct}\relax
\EndOfBibitem
\bibitem[Tieleman \latin{et~al.}(1999)Tieleman, Sansom, and
  Berendsen]{tieleman1999}
Tieleman,~D.~P.; Sansom,~M.~S.; Berendsen,~H.~J. Alamethicin Helices in a
  Bilayer and in Solution: Molecular Dynamics Simulations. \emph{Biophys. J.}
  \textbf{1999}, \emph{76}, 40--49\relax
\mciteBstWouldAddEndPuncttrue
\mciteSetBstMidEndSepPunct{\mcitedefaultmidpunct}
{\mcitedefaultendpunct}{\mcitedefaultseppunct}\relax
\EndOfBibitem
\bibitem[B{\"o}ckmann \latin{et~al.}(2008)B{\"o}ckmann, De~Groot, Kakorin,
  Neumann, and Grubm{\"u}ller]{Bockmann2008kinetics}
B{\"o}ckmann,~R.~A.; De~Groot,~B.~L.; Kakorin,~S.; Neumann,~E.;
  Grubm{\"u}ller,~H. Kinetics, statistics, and energetics of lipid membrane
  electroporation studied by molecular dynamics simulations. \emph{Biophys. J.}
  \textbf{2008}, \emph{95}, 1837--1850\relax
\mciteBstWouldAddEndPuncttrue
\mciteSetBstMidEndSepPunct{\mcitedefaultmidpunct}
{\mcitedefaultendpunct}{\mcitedefaultseppunct}\relax
\EndOfBibitem
\bibitem[Hockney and Eastwood(1989)Hockney, and Eastwood]{hockney1989pppm}
Hockney,~R.; Eastwood,~J. \emph{Computer Simulation Using Particles}; Adam
  Hilger, New York, 1989\relax
\mciteBstWouldAddEndPuncttrue
\mciteSetBstMidEndSepPunct{\mcitedefaultmidpunct}
{\mcitedefaultendpunct}{\mcitedefaultseppunct}\relax
\EndOfBibitem
\bibitem[Pollock and Glosli(1996)Pollock, and Glosli]{pollocl1996pppm}
Pollock,~E.; Glosli,~J. Comments on P3M, FMM, and the Ewald method for large
  periodic Coulombic systems. \emph{Comput. Phys. Commun.} \textbf{1996},
  \emph{95}, 93--110\relax
\mciteBstWouldAddEndPuncttrue
\mciteSetBstMidEndSepPunct{\mcitedefaultmidpunct}
{\mcitedefaultendpunct}{\mcitedefaultseppunct}\relax
\EndOfBibitem
\bibitem[Plimpton(1995)]{lammps}
Plimpton,~S. Fast parallel algorithms for short-range molecular dynamics.
  \emph{J. Comput. Phys.} \textbf{1995}, \emph{117}, 1--19\relax
\mciteBstWouldAddEndPuncttrue
\mciteSetBstMidEndSepPunct{\mcitedefaultmidpunct}
{\mcitedefaultendpunct}{\mcitedefaultseppunct}\relax
\EndOfBibitem
\bibitem[Allen and Tildesley(1991)Allen, and Tildesley]{Allen91}
Allen,~M.~P.; Tildesley,~D.~J. \emph{Computer Simulations of Liquids}; Oxford
  University Press, Oxford, 1991\relax
\mciteBstWouldAddEndPuncttrue
\mciteSetBstMidEndSepPunct{\mcitedefaultmidpunct}
{\mcitedefaultendpunct}{\mcitedefaultseppunct}\relax
\EndOfBibitem
\bibitem[Robbins and M\"u{}ser(2001)Robbins, and M\"u{}ser]{Robbins01}
Robbins,~M.~O.; M\"u{}ser,~M. In \emph{Modern Tribology Handbook}; Bhushan,~B.,
  Ed.; CRC Press, Boca Raton, FL, 2001; pp 717--825\relax
\mciteBstWouldAddEndPuncttrue
\mciteSetBstMidEndSepPunct{\mcitedefaultmidpunct}
{\mcitedefaultendpunct}{\mcitedefaultseppunct}\relax
\EndOfBibitem
\bibitem[Rottler and Robbins(2003)Rottler, and Robbins]{rottler2003growth}
Rottler,~J.; Robbins,~M.~O. Growth, microstructure, and failure of crazes in
  glassy polymers. \emph{Phys. Rev. E} \textbf{2003}, \emph{68}, 011801\relax
\mciteBstWouldAddEndPuncttrue
\mciteSetBstMidEndSepPunct{\mcitedefaultmidpunct}
{\mcitedefaultendpunct}{\mcitedefaultseppunct}\relax
\EndOfBibitem
\bibitem[Galuschko \latin{et~al.}(2010)Galuschko, Spirin, Kreer, Johner,
  Pastorino, Wittmer, and Baschnagel]{galuschko2010}
Galuschko,~A.; Spirin,~L.; Kreer,~T.; Johner,~A.; Pastorino,~C.; Wittmer,~J.;
  Baschnagel,~J. Frictional forces between strongly compressed, nonentangled
  polymer brushes: molecular dynamics simulations and scaling theory.
  \emph{Langmuir} \textbf{2010}, \emph{26}, 6418--6429\relax
\mciteBstWouldAddEndPuncttrue
\mciteSetBstMidEndSepPunct{\mcitedefaultmidpunct}
{\mcitedefaultendpunct}{\mcitedefaultseppunct}\relax
\EndOfBibitem
\bibitem[Dong \latin{et~al.}(2011)Dong, Perez, Voter, and Martini]{Dong11}
Dong,~Y.; Perez,~D.; Voter,~A.; Martini,~A. The roles of statics and dynamics
  in determining transitions between atomic friction regimes. \emph{Tribol.
  Lett.} \textbf{2011}, \emph{42}, 99--107\relax
\mciteBstWouldAddEndPuncttrue
\mciteSetBstMidEndSepPunct{\mcitedefaultmidpunct}
{\mcitedefaultendpunct}{\mcitedefaultseppunct}\relax
\EndOfBibitem
\bibitem[Drummond and Israelachvili(2001)Drummond, and
  Israelachvili]{Drummond01}
Drummond,~C.; Israelachvili,~J. Dynamic phase transitions in confined lubricant
  fluids under shear. \emph{Phys. Rev. E} \textbf{2001}, \emph{63},
  041506\relax
\mciteBstWouldAddEndPuncttrue
\mciteSetBstMidEndSepPunct{\mcitedefaultmidpunct}
{\mcitedefaultendpunct}{\mcitedefaultseppunct}\relax
\EndOfBibitem
\bibitem[Taylor(1997)]{taylor1997introduction}
Taylor,~J. \emph{Introduction to error analysis, the study of uncertainties in
  physical measurements}; University Science Books, New York, 1997\relax
\mciteBstWouldAddEndPuncttrue
\mciteSetBstMidEndSepPunct{\mcitedefaultmidpunct}
{\mcitedefaultendpunct}{\mcitedefaultseppunct}\relax
\EndOfBibitem
\bibitem[Filippov \latin{et~al.}(2004)Filippov, Klafter, and
  Urbakh]{Filippov04}
Filippov,~A.; Klafter,~J.; Urbakh,~M. Friction through dynamical formation and
  rupture of molecular bonds. \emph{Phys. Rev. Lett.} \textbf{2004}, \emph{92},
  135503\relax
\mciteBstWouldAddEndPuncttrue
\mciteSetBstMidEndSepPunct{\mcitedefaultmidpunct}
{\mcitedefaultendpunct}{\mcitedefaultseppunct}\relax
\EndOfBibitem
\bibitem[Guerra \latin{et~al.}(2016)Guerra, Benassi, Vanossi, Ma, and
  Urbakh]{Guerra2016}
Guerra,~R.; Benassi,~A.; Vanossi,~A.; Ma,~M.; Urbakh,~M. Friction and adhesion
  mediated by supramolecular host--guest complexes. \emph{Phys. Chem. Chem.
  Phys.} \textbf{2016}, \emph{18}, 9248--9254\relax
\mciteBstWouldAddEndPuncttrue
\mciteSetBstMidEndSepPunct{\mcitedefaultmidpunct}
{\mcitedefaultendpunct}{\mcitedefaultseppunct}\relax
\EndOfBibitem
\bibitem[Blass \latin{et~al.}(2017)Blass, Albrecht, Wenz, Guerra, Urbakh, and
  Bennewitz]{Blass2017}
Blass,~J.; Albrecht,~M.; Wenz,~G.; Guerra,~R.; Urbakh,~M.; Bennewitz,~R.
  Multivalent adhesion and friction dynamics depend on attachment flexibility.
  \emph{J. Phys. Chem. C} \textbf{2017}, \emph{121}, 15888--15896\relax
\mciteBstWouldAddEndPuncttrue
\mciteSetBstMidEndSepPunct{\mcitedefaultmidpunct}
{\mcitedefaultendpunct}{\mcitedefaultseppunct}\relax
\EndOfBibitem
\bibitem[Liu and Szlufarska(2012)Liu, and Szlufarska]{Liu12a}
Liu,~Y.; Szlufarska,~I. Chemical origins of frictional aging. \emph{Phys. Rev.
  Lett.} \textbf{2012}, \emph{109}, 186102\relax
\mciteBstWouldAddEndPuncttrue
\mciteSetBstMidEndSepPunct{\mcitedefaultmidpunct}
{\mcitedefaultendpunct}{\mcitedefaultseppunct}\relax
\EndOfBibitem
\bibitem[Li \latin{et~al.}(2011)Li, Tullis, Goldsby, and Carpick]{Li11b}
Li,~Q.; Tullis,~T.; Goldsby,~D.; Carpick,~R. Frictional ageing from interfacial
  bonding and the origins of rate and state friction. \emph{Nature}
  \textbf{2011}, \emph{480}, 233--236\relax
\mciteBstWouldAddEndPuncttrue
\mciteSetBstMidEndSepPunct{\mcitedefaultmidpunct}
{\mcitedefaultendpunct}{\mcitedefaultseppunct}\relax
\EndOfBibitem
\bibitem[Tian \latin{et~al.}(2017)Tian, Gosvami, Goldsby, Liu, Szlufarska, and
  Carpick]{Tian2017}
Tian,~K.; Gosvami,~N.~N.; Goldsby,~D.~L.; Liu,~Y.; Szlufarska,~I.;
  Carpick,~R.~W. Load and time dependence of interfacial chemical bond-induced
  friction at the nanoscale. \emph{Phys. Rev. Lett.} \textbf{2017}, \emph{118},
  076103\relax
\mciteBstWouldAddEndPuncttrue
\mciteSetBstMidEndSepPunct{\mcitedefaultmidpunct}
{\mcitedefaultendpunct}{\mcitedefaultseppunct}\relax
\EndOfBibitem
\bibitem[Ouyang \latin{et~al.}(2019)Ouyang, Ramakrishna, Rossi, Urbakh,
  Spencer, and Arcifa]{Ouyang2019}
Ouyang,~W.; Ramakrishna,~S.~N.; Rossi,~A.; Urbakh,~M.; Spencer,~N.~D.;
  Arcifa,~A. Load and velocity dependence of friction mediated by dynamics of
  interfacial contacts. \emph{Phys. Rev. Lett.} \textbf{2019}, \emph{123},
  116102\relax
\mciteBstWouldAddEndPuncttrue
\mciteSetBstMidEndSepPunct{\mcitedefaultmidpunct}
{\mcitedefaultendpunct}{\mcitedefaultseppunct}\relax
\EndOfBibitem
\bibitem[Shao \latin{et~al.}(2017)Shao, Jacobs, Jiang, Turner, Carpick, and
  Falk]{Shao2017}
Shao,~Y.; Jacobs,~T.~D.; Jiang,~Y.; Turner,~K.~T.; Carpick,~R.~W.; Falk,~M.~L.
  Multibond model of single-asperity tribochemical wear at the nanoscale.
  \emph{ACS Appl. Mater. Inter.} \textbf{2017}, \emph{9}, 35333--35340\relax
\mciteBstWouldAddEndPuncttrue
\mciteSetBstMidEndSepPunct{\mcitedefaultmidpunct}
{\mcitedefaultendpunct}{\mcitedefaultseppunct}\relax
\EndOfBibitem
\bibitem[Karuppiah \latin{et~al.}(2009)Karuppiah, Zhou, Woo, and
  Sundararajan]{karuppiah2009}
Karuppiah,~K.~K.; Zhou,~Y.; Woo,~L.~K.; Sundararajan,~S. Nanoscale friction
  switches: friction modulation of monomolecular assemblies using external
  electric fields. \emph{Langmuir} \textbf{2009}, \emph{25}, 12114--12119\relax
\mciteBstWouldAddEndPuncttrue
\mciteSetBstMidEndSepPunct{\mcitedefaultmidpunct}
{\mcitedefaultendpunct}{\mcitedefaultseppunct}\relax
\EndOfBibitem
\end{mcitethebibliography}

\providecommand{\latin}[1]{#1}
\makeatletter
\providecommand{\doi}
  {\begingroup\let\do\@makeother\dospecials
  \catcode`\{=1 \catcode`\}=2 \doi@aux}
\providecommand{\doi@aux}[1]{\endgroup\texttt{#1}}
\makeatother
\providecommand*\mcitethebibliography{\thebibliography}
\csname @ifundefined\endcsname{endmcitethebibliography}
  {\let\endmcitethebibliography\endthebibliography}{}

\end{document}

\newpage
\section{Supercell Geometry}

Intending to arrange periodically two arrays of molecules according to triangular lattices we introduce primitive vectors 
\begin{align}
    \mathbf{a}_1&=a \left(\cos (\phi/2),\sin(\phi/2)\right)\qquad
    &\mathbf{a}_2=a \left(\cos (\pi/3+\phi/2),\sin(\pi/3+\phi/2)\right)\\
    \mathbf{b}_1&=a \left(\cos (-\phi/2),\sin(-\phi/2)\right)\qquad
    &\mathbf{b}_2=a \left(\cos (\pi/3-\phi/2),\sin(\pi/3-\phi/2)\right)
\end{align}
with spacing $a = 0.82$~nm, and rotated by $\phi /2$ in opposite directions.

We need a common periodicity and therefore, a matching lattice vector
%13*Sin(x)+7*Sin(π/3+x)==-7*Sin(x)+13*Sin(π/3-x)
\begin{align}\label{eq:matching}
    m_1 \mathbf{a}_1 + m_2 \mathbf{a}_2 = m_2 \mathbf{b}_1 + m_1 \mathbf{b}_2
\end{align}
with integers $m_1$ and $m_2$. This evaluation of this common lattice vector is a special case of the theory described in Ref.~\onlinecite{Grey92}.

We find that a reasonably-sized supercell is obtained with $m_1=13$ and $m_2=7$. We solve, e.g., the $y$ component of Eq.~\eqref{eq:matching}:
\begin{align}
    m_1\sin(\phi/2)+m_2\sin(\pi/3+\phi/2)=m_1\sin(\pi/3-\phi/2)-m_2\sin(\phi/2).
\end{align}
We obtain
\begin{align}
    \phi = 2 \arctan \left(\frac{\sqrt{3}}{10}\right) \simeq 19.65286^\circ.
\end{align}

With the adopted geometry both rotated lattices have lattice points along $x$ and $y$. To determine them we solve the null components of the equations
\begin{align}\label{eq:matchingN}
    n_1 \mathbf{a}_1 + n_2 \mathbf{a}_2 &= (l_x,0) \\
    n'_1 \mathbf{a}_1 + n'_2 \mathbf{a}_2 &= (0,l_y), 
\end{align}
obtaining $n_1=11$, $n_2=-2$, $n'_1 =-7$ and $n'_2=20$. 
The remaining components of the equations provide 
\begin{align}
    l_x&=a \left[n_1\cos (\phi/2) +n_2 \cos (\phi/2+\pi/3)\right] \simeq 8.32~\text{nm}\\
    l_y&=a \left[n'_1\sin (\phi/2) +n'_2 \sin (\phi/2+\pi/3)\right] \simeq 14.41~\text{nm}. 
\end{align}
Alternatively, the opposite corner $(l_x,l_y)$ of a rectangular supercell with a corner at the origin $(0,0)$ is obtained as $(l_x,l_y) = 4\mathbf{a}_1+18\mathbf{a}_2$.
%\clearpage
%\newpage

\section{Cutoff of the Two-Body Potential}
For the non-bonded pairwise particle-particle interactions we adopt a Morse potential with a standard shift and a linear term added as follows:
\begin{equation}
    V\left(r \right) 
    = 
    \left\{\begin{array}{lc}
         V_\text{Morse}\left(r \right)-V_\text{Morse}\left(R_c \right)-\left( r-R_c \right) \left. \frac{dV_\text{Morse}}{dr} \right|_{r=R_c},& r<R_c \\
         0, &  r\geq R_c
    \end{array}\right.
\end{equation}
so that the truncated potential vanishes smoothly at $R_c$.

\section{Long-Range Solver for Coulomb Interactions}

The PPPM solver used for systems, such as ours, which are periodic in $x$ and $y$, but not in $z$, requires an ad-hoc extension.
The system is treated as if it was periodic in $z$, but inserting an empty volume between the slabs
and thus removing unphysical dipole inter-slab interactions.
For the parameter setting the fraction of empty volume in between slab repetitions, we adopt the value $3.0$ recommended by the developers of the simulation software LAMMPS \cite{yeh1999ewald}.
We explicitly verified that, by improving the accuracy of the PPPM solver beyond $10^{-4}$\,eV/nm, neither quantitative effects on the sliding friction nor qualitative effects on the system dynamics are detectable.

\clearpage\newpage

\begin{figure}[tb]
\includegraphics[width=0.7\textwidth]{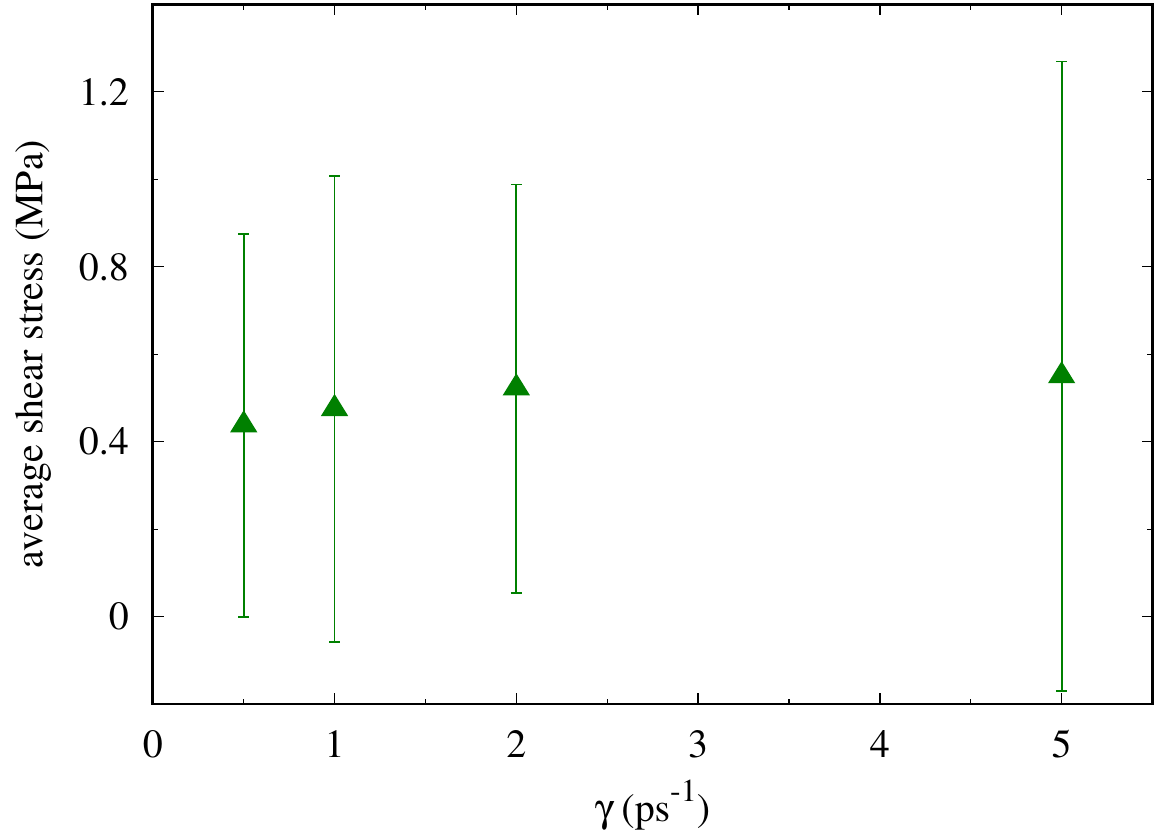}
\caption{
    \label{damping} The average shear stress as a function of the damping parameter $\gamma$ for the %charged 
    zwitterionic
    model. Simulations are carried out for $v_{\text stage} = 2.5$\,m$\cdot$s$^{-1}$,  $T = 150$\,K, $L = 10$\,MPa.}
\end{figure}

\clearpage\newpage

\section{The Hooking Fraction $h$}
In order to quantify the degree of interpenetration of the chains we introduce a ``hooking fraction'' $h$ as the fractional number of chains whose cation crosses the average level of cations of the opposite layer, like the highlighted chains in Fig.~4b,d of the main text. 
The definition for $h$ is the following:
\begin{equation}
	h=\frac{1}{2}\left( h_{\rm SUP}+h_{\rm SUB}\right),
\end{equation}
where 
\begin{align}
    h_{\rm SUB} = \frac{1}{N_{\rm SUB}}\sum _{j=1}^{N_{\rm SUB}}\theta (z^{\rm CA}_j - \bar z_{\rm SUP})\\
        h_{\rm SUP} = \frac{1}{N_{\rm SUP}}\sum _{i=1}^{N_{\rm SUP}}\theta (\bar z_{\rm SUB} - z^{\rm CA}_i  ).
\end{align}
Here $\theta()$ is the usual $\theta$ function, equal to one or zero according to the sign of its argument, and
\begin{equation}
    \bar z_{\rm SUP} = \frac{1}{N_{\rm SUP}}\sum _{i=1}^{N_{\rm SUP}}z^{\rm CA}_i
    ,\qquad 
    \bar z_{\rm SUB} = \frac{1}{N_{\rm SUB}}\sum _{j=1}^{N_{\rm SUB}}z^{\rm CA}_j.
\end{equation}

For example, the hooking fraction as a function of time is illustrated in Fig.~5a. 
$h$ clearly correlates with the stick-slip dynamics.

\newpage

\begin{figure}[tb]
\includegraphics[width=\textwidth]{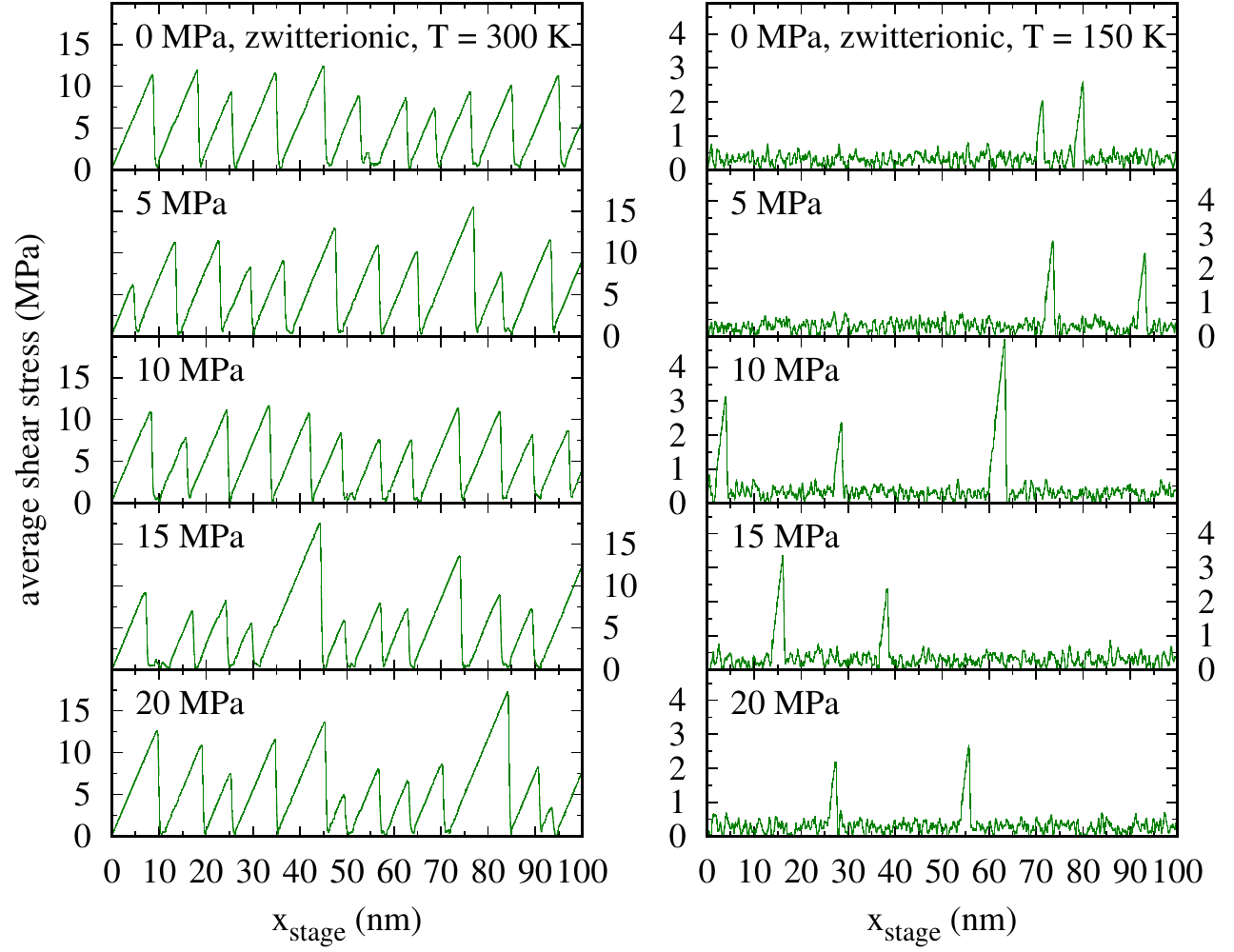}
\caption{
    \label{shear-load-charged} The frictional shear traces of the %charged
    zwitterionic
    model obtained for stepwise increasing load $L$. Averages (including the load-decreasing traces -- not shown) are reported in Fig.~7a of the main text. Simulation conditions: $v_{\text stage} = 5$\,m$\cdot$s$^{-1}$, (left) $T = 300$\,K and (right) $T = 150$\,K.}
\end{figure}

\clearpage\newpage 

\begin{figure}[tb]
\includegraphics[width=\textwidth]{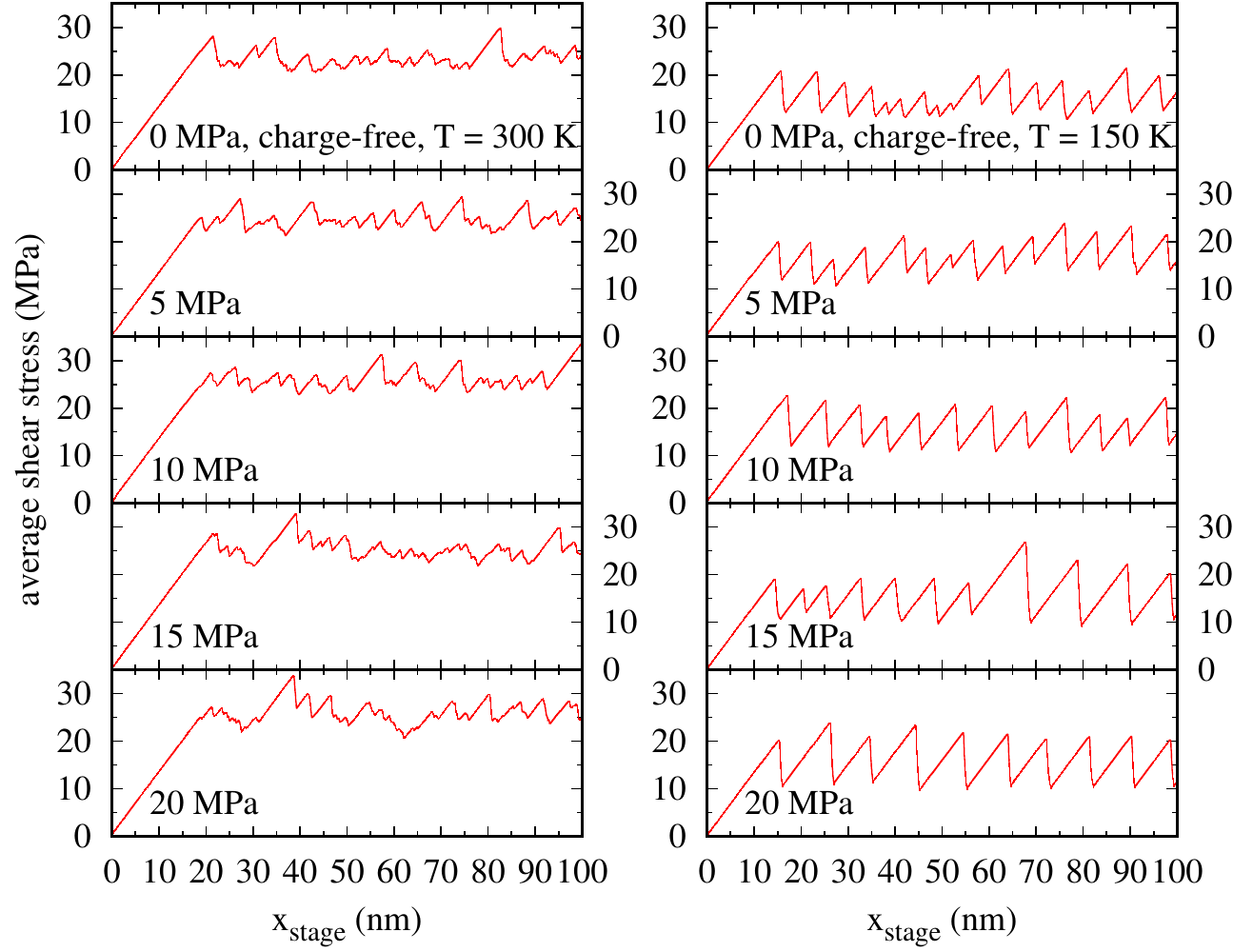}
\caption{
    \label{shear-load-uncharged} Same as Fig.~\ref{shear-load-charged} but for the %uncharged 
    charge-free
    model -- Averages reported in Fig.~7c of the main text.}
\end{figure}

\newpage

\begin{figure}[tb]
\includegraphics[width=0.8\textwidth]{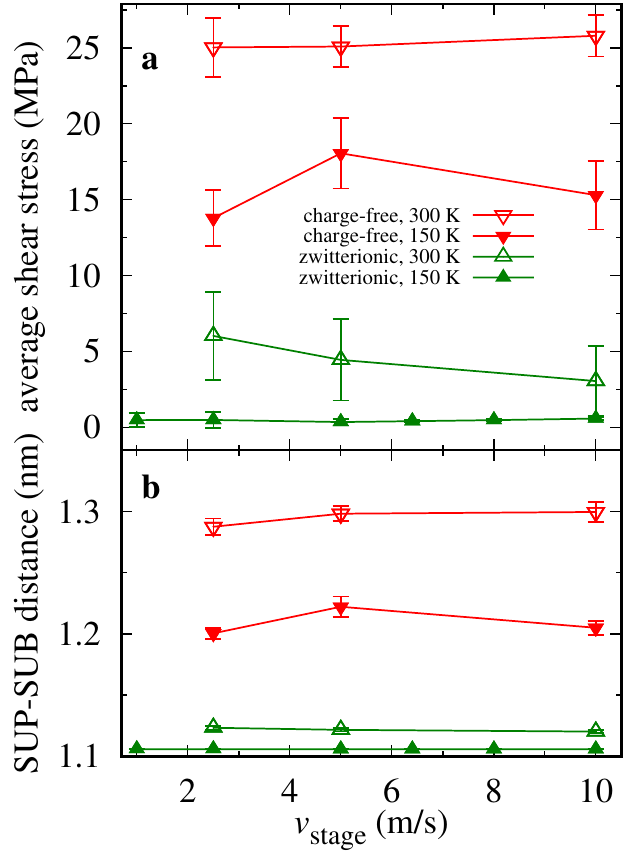}
\caption{
    \label{fig:vX} (a) The frictional shear stress and 
    (b) the distance between the rigid layers as a function of the advancement velocity of the stage for the two indicated temperatures and for load $L = 10$\,MPa.}
\end{figure}

\clearpage\newpage

\begin{figure}[tb]
\includegraphics[width=\textwidth]{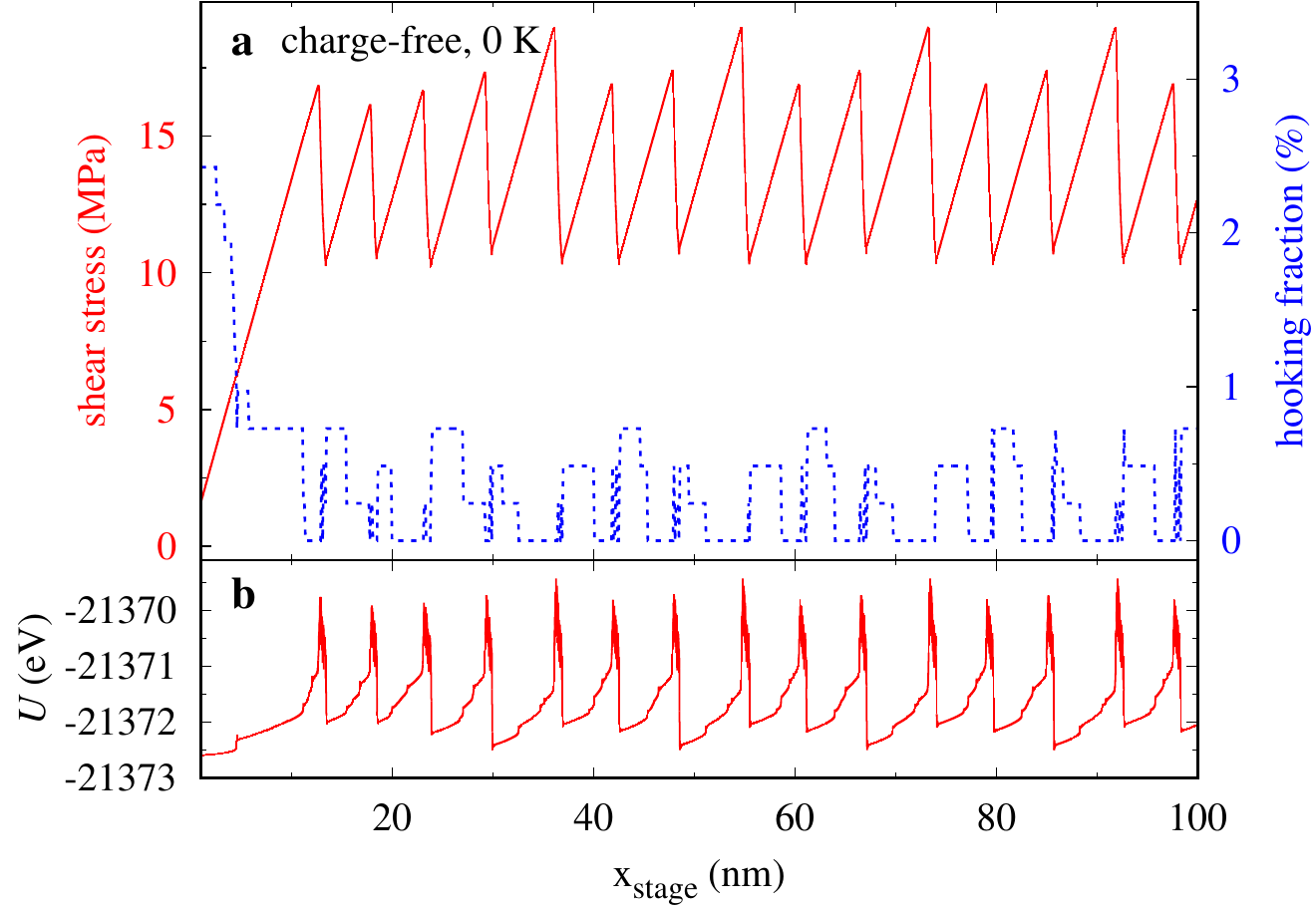}
\caption{
    \label{shear-h_unharged} (a) The percentile hooking fraction $h$ as a function of the stage displacement correlated with the frictional shear stress for the %uncharged 
    charge-free
    model at $T=0$~K. (b) The total potential energy for the same simulation.}
\end{figure}

\clearpage\newpage

\begin{figure}[tb]
\includegraphics[width=\textwidth]{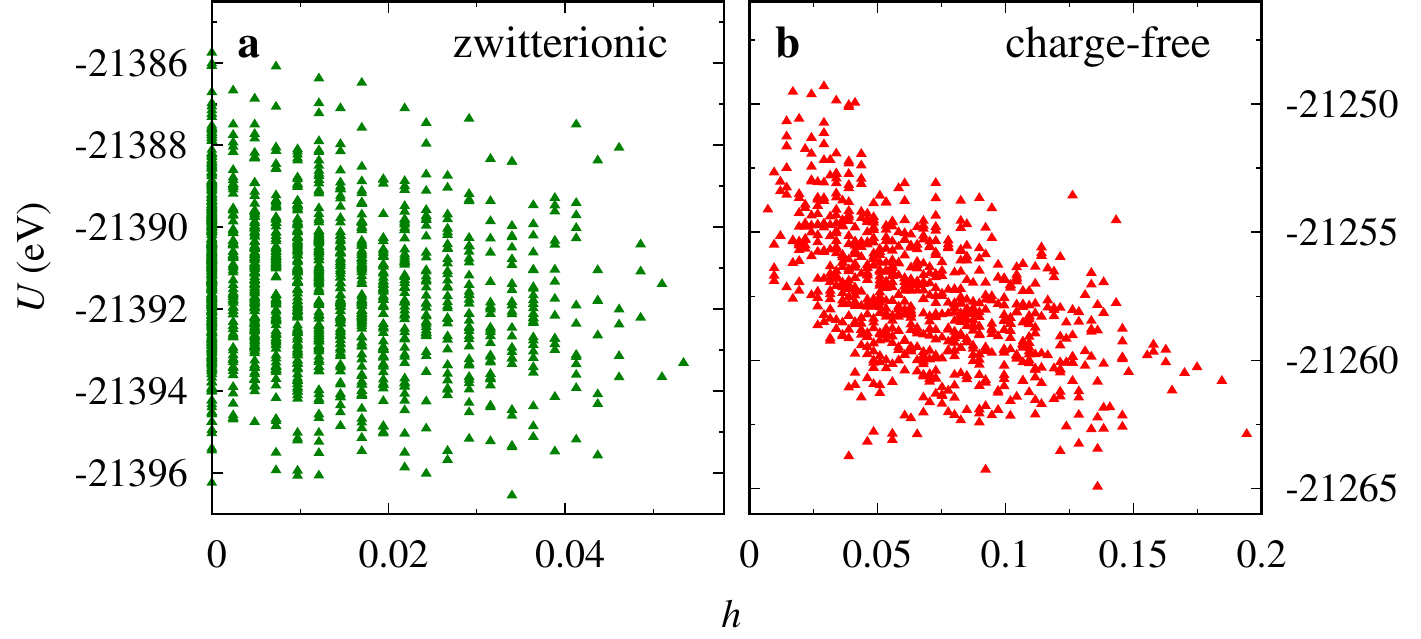}
\caption{\label{corr-300} Scatter plot illustrating the correlation between the total potential energy and the hooked fraction for (a) %charged 
zwitterionic
system (b) %uncharged 
charge-free
system. Correlation coefficients for these data are reported in Figure 8b of the paper. $L =10$~MPa and $T=300$~K.}
\end{figure}

\clearpage\newpage

\section{SI Movies}

Each of the SI movies reports the final 6~ns (i.e. \-the last 30~nm displacement) of a MD simulation.
In simulation time, the frame rate is 1 frame every 20~ps. In running time, the frame rate is 10 frames per second. For clarity, like in Fig.~3 of the main text, the movies only include a 5 nm $y$-thick slice of the simulation cell (whose entire $y$-side is 14.41~nm).

Each movie contains one highlighted SUP particle to make the displacement of the rigid top layer more evident.
\begin{itemize}
    \item \textbf{zwitterionic\_150K.mp4}: the last 6~ns of the MD simulation corresponding to the force trace shown in Figure 2c;
    \item \textbf{zwitterionic\_300K.mp4}: the last 6~ns of the MD simulation corresponding to the force trace shown in Figure 2d; 
    \item \textbf{charge-free\_150K.mp4}: the last 6~ns of the MD simulation corresponding to the force trace shown in Figure 2e;
    \item \textbf{charge-free\_300K.mp4}: the last 6~ns of the MD simulation corresponding to the force trace shown in Figure 2f. 
\end{itemize}

%\newpage

%\bibliography{biblioMMG.bib,biblio.bib}

\providecommand{\latin}[1]{#1}
\makeatletter
\providecommand{\doi}
  {\begingroup\let\do\@makeother\dospecials
  \catcode`\{=1 \catcode`\}=2 \doi@aux}
\providecommand{\doi@aux}[1]{\endgroup\texttt{#1}}
\makeatother
\providecommand*\mcitethebibliography{\thebibliography}
\csname @ifundefined\endcsname{endmcitethebibliography}
  {\let\endmcitethebibliography\endthebibliography}{}